\documentclass[11pt]{raa}            % referee version: for submission

\usepackage{graphicx,times}             %for PS/EPS graphics inclusion, new
\usepackage{natbib}
\usepackage{amssymb,amsmath}
\usepackage{epstopdf} % eps转PDF
\usepackage{multirow}
\usepackage{multicol}
\usepackage{longtable}
\usepackage{lscape}
\bibpunct{(}{)}{;}{a}{}{,}

\usepackage[a4paper,top=25mm,left=25mm,bottom=20mm,centering]{geometry}
\usepackage[pagebackref=true]{hyperref}
\hypersetup{colorlinks = true, linkcolor = green, anchorcolor = red, citecolor = blue, filecolor = red, pagecolor = red, urlcolor = red}

\begin{document}

\title{A Catalog of Molecular Clumps and Cores with Infall Signatures
   %\footnotetext{$*$ Supported by the National Natural Science Foundation of China.}
}
%   \subtitle{I. Place Your Subtitle Here}

\volnopage{Vol.0 (20xx) No.0, 000--000}      %%preserved for Editor. DOn't remove!
\setcounter{page}{1}          %%starting page, preserved for Editor. DOn't remove!

\author{Shuling Yu
   \inst{1,2}
   \and Zhibo Jiang
   \inst{1,2,3}
   \and Yang Yang
   \inst{1,2}
   \and Zhiwei Chen
   \inst{1}
   \and Haoran Feng
   \inst{1,2}
}

\institute{Purple Mountain Observatory, Chinese Academy of Sciences, Nanjing, Jiangsu 210023, People's Republic of China; {\it zbjiang@pmo.ac.cn}\\
\and
University of Science and Technology of China, Chinese Academy of Sciences, Hefei, Anhui 230026, People's Republic of China\\
\and
China Three Gorges University, Yichang, Hubei 443002, People's Republic of China\\
\vs\no
{\small Received 20xx month day; accepted 20xx month day}}

\abstract{ The research of infall motion is a common means to study molecular cloud dynamics and the early process of star formation. Many works had been done in-depth research on infall. We searched the literature related to infall study of molecular cloud since 1994, summarized the infall sources identified by the authors. A total of 456 infall sources are catalogued. We classify them into high-mass and low-mass sources, in which the high-mass sources are divided into three evolutionary stages: prestellar, protostellar and {H\small \hspace{0.1em}II} region. 
 We divide the sources into clumps and cores according to their sizes.
 The H$_2$ column density values range from 1.21$\times$ 10$^{21}$ to 9.75 $\times$ 10$^{24}$ cm$^{-2}$, with a median value of 4.17$\times$ 10$^{22}$ cm$^{-2}$. The H$_2$ column densities of high-mass and low-mass sources are significantly separated.
 The median value of infall velocity for high-mass clumps is 1.12 km s$^{-1}$, and the infall velocities of low-mass cores are virtually all less than 0.5  km s$^{-1}$. There is no obvious difference between different stages of evolution.
 The mass infall rates of low-mass cores are between 10$^{-7}$ and 10$^{-4}$ M$_{\odot} \text{yr}^{-1}$, and those of high-mass clumps are between 10$^{-4}$ and  10$^{-1}$ M$_{\odot} \text{yr}^{-1}$ with only one exception. We do not find that the mass infall rates vary with evolutionary stages.
 \keywords{stars: formation --- ISM: kinematics and dynamics --- ISM: molecules --- radio lines: ISM}
}

\authorrunning{\it Yu et al.}            %author_head in even pages
\titlerunning{\it A Catalog of Infall Signatures}             % title_head in odd pages

\maketitle
%% The author head (on even pages) and the title head (on odd pages) will be
%% automatically extracted from \author{} and \title{}. Whenever the title is too long,
%% you will be asked to supply a shorter one by inserting either \authorrunning{} or
%% \titlerunning{} before \maketitle. Anyway, you can specify your own heads.
%%
%%
%% Note: In the following text body of your manuscript, please note several differences from
%%       other major journals:
%% (1) \subsection{Please Capitalize the First Letter of Each Notional Word in Subsection Title}
%% (2) Please Capitalize the First Letter of Each Notional Word in all tables' captions

%
%________________________________________________ sections below
% 
\section{Introduction}           %% first-level sections will be auto-capitalized
\label{sect:intro}

Infall motion is the initial stage of star formation. Although stars with different masses may have different formation mechanisms (e.g., \citealt{Cesaroni2007,Zinnecker2007,Beltran2016,Klassen2016}), they all start from the collapse and infall of molecular cloud cores.
Low-mass stars are believed to formed through inside-out gravitational collapse of dense molecular cores (\citealt{Shu1987}), in which gravity overcomes thermal and nonthermal (mainly magnetic and turbulent) pressure (\citealt{Zhou1993,Zhou1994}).
For high-mass stars ($\geqslant $8 $M_\odot$), the formation mechanism is still controversial because of the poverty of samples and the difficulty of observation due to the larger distance of star-forming region, stronger interaction with surrounding molecular clouds, and shorter star-forming timescale.
There are two dominant models, one is core accretion (\citealt{Yorke2002,McKee2002,McKee2003}), which assumes that massive stars form as their population of low-mass stars in isolation, but that the accretion rate is a few orders of magnitude higher than low-mass stars.
Another model is called competitive accretion (\citealt{Bonnell2002,Bonnell2004}). This model claims that massive stars form in the dense part of molecular clumps (size $\leqslant $1 pc and mass $\geqslant $200 $M_\odot$, \citealt{Beuther2005,Tan2014}).
They compete to accrete materials with other protostars, and the accretion continues until the accretion condition is not satisfied. Some recent observations (e.g., \citealp*{Liu2013,Peretto2013,Contreras2018}) support this model.
In any case, infall motion is the key to initiating the formation of massive stars and maintaining accretion flows to increase mass continually during subsequent phases of evolution in these two models.

The method of finding infall samples is to identify them through special line profile. One of the significant features of infall motion is called ``the spectral line asymmetry" (e.g., \citealt{Zhou1992,Mardones1997,Lee1999,Evans1999,Wu2003,Churchwell2010}), i.e. it shows double peaks profile where the blue peak is stronger than the red one on an optically thick line, while the optically thin line has only a single peak between the double peaks of the optically thick line.
The commonly used optically thick lines are CO(1-0), HCO$^+$(1-0), HCO$^+$(4-3), HCN(1-0), CS(2-1), H$_2$CO(2-1), etc.
Their isotopic molecular line, C$^{18}$O(1-0), H$^{13}$CO$^+$(1-0), H$^{13}$CO$^+$(4-3), H$^{13}$CO$^+$(3-2), H$^{13}$CN(1-0), are generally used as optically thin line to trace the system velocity.
In addition, N$_2$H$^+$(1-0) is generally optically thin, and can characterize the dense inner region of a core and well estimate the system velocity (\citealt{Mardones1997,Lee1999}), so it is particularly suitable for studying the structure and kinematics of star-forming cores and has been used to detect infall in IRDCs (infrared dark clouds), starless cores and protostar cores (e.g., \citealt{Lee2001,Fuller2005,Keto2015,Keown2016,Storm2016,Kim2021}).

At present, there are good constraints on the formation of low-mass stars, and the corresponding evidence of infall motion has been found in the early stage until class I phase  (e.g., \citealt{Mardones1997,Gregersen2000,Lee2001,Lee2011,Padoan2014,Keown2016,Kim2021}). In contrast, due to the controversy on the formation mechanism, the infall observation and research of high-mass young stellar objects (YSOs) has become more popular in recent years (e.g., \citealt{Sridharan2002,Fuller2005,Klaassen2008,Klaassen2012,Sun2009,Reiter2011,Liu2011,Qiu2012,Liu2013,He2016,Qin2016,Cunningham2018,Saral2018,Tang2019,Liu2020,Yang2020,Yue2021}).
\citet{Rygl2013} found that the formation are closely related to the ``clump-to-cloud'' column density ratio. They did not find YSOs in clumps with molecular hydrogen column density less than 4 $\times$ 10$^{22}$ cm$^{-2}$; this ratio greater than 2 is the first sign of the beginning of star formation, and infall signature can be widely observed at this time; when the ratio greater than 3, the infall motions stop in most clumps.
\citet{He2015,He2016} analysed 732 high-mass clumps, and concluded that the detection rate of infall sources decreases gradually from prestellar to protostar and then to {H\small \hspace{0.1em}II} region phase; the mass of the clumps increase with these evolutionary stages whether the infall is detected or not.
However, \citet{Klaassen2012} came to an abnormal situation that the detection rate of infall candidates increased significantly from protostar to hypercompact {H\small \hspace{0.1em}II} region phase. He thought this may be related to the beam dilution caused by smaller infall area of younger sources.
\citet{Yue2021} observed 30 high-mass clumps associated with bright infrared sources, and found that the clumps with higher mass and luminosity tend to have larger mass infall rates.
Large sample survey and high resolution observation of interferometric array have become the future trend of infall study.

On the basis of observation and theoretical study, the process of massive star formation is proposed by a number of researchers. \citet{Saral2018} used the classification method summarized by \citet{Motte2018}: (1) starless core (\citealt{Lee2001,Sohn2007,Lee2011,Schnee2013}), the gravitationally bound prestellar core or IRDC (\citealt{Contreras2018}); (2) hot molecular core (HMC) or protostar (\citealt{Mardones1997,Wu2005,Fuller2005,Keown2016}); (3) {H\small{II}} region (\citealt{He2016}).
An additional stage is defined by \citet{Guzman2015}. At this stage, the expansion of the ionized gas finally destroys the molecular envelope, forming the characteristic of an extended classical {H\small \hspace{0.1em}II} region and a photodissociation region (PDR). Because {H\small \hspace{0.1em}II} still exists in this stage, but it is merely more extended, so we incorporate this stage into the stage 3.
Therefore, this paper adopts the classification by \citet{Motte2018}, which is divided into three main stages: prestellar, protostellar, {H\small \hspace{0.1em}II} region.

For simplicity, we define the infall source as source with infall signature in this paper. There is no complete statistical work on infall sources before. Therefore we make a comprehensive search of infall study in the literature, aggregate the infall sources into a catalog, and make some statistical analysis on the physical properties.

\section{The Data}           %% first-level sections will be auto-capitalized
\label{sect:The Catalogue}

\subsection{Tracers}

Table~\ref{Tab:spectral-line} lists the optically thick and thin lines used to identify the infall signatures collected in this paper. The frequencies and upper level of energies ($E_{\text{u}}$) in the catalogue are obtained from the Splatalogue Database\footnote{\url{https://splatalogue.online}}. The critical density ($n_{\text{crit}}$) of each molecular transition is calculated by $n_{\text{crit}}=A_{u l} / \gamma_{ul}$,
where $A_{u l}$ is the Einstein A-coefficient and $\gamma_{u l}$ is the collision rate coefficient for collisions with H$_2$ at a certain temperature.
The Leiden Atomic and Molecular Database\footnote{\url{https://home.strw.leidenuniv.nl/~moldata/}} (LAMDA, \citealt{Schoier2005}) provide $A_{u l}$ and $\gamma_{u l}$ at different temperatures. We adopt those which are nearest $E_\text{u}$.

Different spectral lines can trace diffrent properties.
HCO$^+$ is an abundant molecule, with particularly enhanced abundances around higher fractional ionized regions (\citealt{Vasyunina2011}).
HCN is a good tracer of infall motion in the low-mass star-forming region. However, it may become unreliable because of a higher level of turbulence (\citealt{Redman2008}) and outflow signatures (\citealt{Zhang2007}) for the high-mass cores (\citealt{Vasyunina2011}).
HNC is also a commonly used infall tracer. The abundance ratio HCN/HNC strongly depends on the temperature, it decreases from decades near the warm core to $\sim$1 on the colder edges (\citealt{Sarrasin2010}).

Table~\ref{Tab:sources-lines} list the number of sources observed by each spectral line. The upper part is the optically thick line.
About 77\% of the infall sources are identified with HCO$^+$ (1-0), HNC (1-0) is the second most (28\%).
This may because HCO$^+$ (1-0) and HNC (1-0) have moderate critical density, which are high enough to trace the dense part closer to the core, and are not as high as CS (2-1), which makes it difficult to achieve such environmental conditions.
In addition, the frequency of HCO$^+$(1-0) and HNC (1-0) fall in one of the best windows of atmospheric transparency, allowing them to be easily observed.
\citet{Smith2013} modelled the emission produced by the hydrodynamic simulation of collapsing clouds in massive star-forming regions without outflows, and concluded that HCO$^+$ (1-0) is better performed to trace collapse than HCO$^+$ (4-3).
Some infall studies (e.g., \citealt{Fuller2005,Vasyunina2011,Rygl2013,Cunningham2018,Yuan2018,Zhang2018}) that simultaneously observe multiple optically thick lines seem to support HCO$^+$ (1-0) as a more effective tracer.
But note that this may not be always true, for example, \citet{Tsamis2008} shows that HCO$^+$ (4-3) may be a better tracer to detect the asymmetry than HCO$^+$ (1-0) at earlier times.

The observation of optically thin lines is also particularly crucial.
Multiple velocity component, for example, the colliding fragments of two clouds, could produce the double-peaked blue profile, but they would also produce a double-peaked profile in the optically thin line (\citealt{Evans1999}).
In order to eliminate this situation, the optically thin line must have only a single peak between the two peaks of the optically thick line.
The lower part of table~\ref{Tab:sources-lines} list the optically thin lines used in the articles.
Different line widths of optically thin lines will affect the estimation of skewness parameters, see section~\ref{subsec:Line profile}.

% % one-column-spanning table
\begin{table}[htbp]
   \begin{center}
      \caption[]{Tracers Used in the Literature}\label{Tab:spectral-line}
      \resizebox{1\textwidth}{!}{   % 自动调整表格宽度 
         %%Please Capitalize the First Letter of Each Notional Word in table's caption

         \begin{tabular}{ccccccc}
            \hline\noalign{\smallskip}
            Serial Number                          &
            Spectral line                          &
            Transition                             &
            Rest Frequency$^{\textcolor{blue}{a}}$ &
            ${E_{\text{u}}}^{\textcolor{blue}{a}}$ &
            ${n_{\text{crit}}}^{\textcolor{blue}{b}}$                                                                                                                  \\
                                                   &                &                               & \scriptsize{(GHz)} & \scriptsize{(K)} & \scriptsize{(cm$^{-3}$)} \\
            \hline\noalign{\smallskip}
            \multicolumn{6}{c}{optically thick lines}                                                                                                                  \\
            \hline
            1                                      & HCO$^+$        & J = 1-0                       & 89.188526          & 4.28             & 1.6 $\times$ 10$^5$      \\
            2                                      & HCO$^+$        & J = 3-2                       & 267.557633         & 25.68            & 3.7 $\times$ 10$^6$      \\
            3                                      & HCO$^+$        & J = 4-3                       & 356.734242         & 42.80            & 9.3 $\times$ 10$^6$      \\
            4                                      & CO             & J = 2-1                       & 230.538000         & 16.60            & 1.1 $\times$ 10$^4$      \\
            5                                      & CO             & J = 3-2                       & 345.795990         & 33.19            & 3.7 $\times$ 10$^4$      \\
            6                                      & CO             & J = 4-3                       & 461.040768         & 55.32            & 9.1 $\times$ 10$^4$      \\
            7                                      & HNC            & J = 1-0                       & 90.663564          & 4.35             & 2.8 $\times$ 10$^5$      \\
            8                                      & HNC            & J = 4-3                       & 362.630304         & 43.51            & 2.0 $\times$ 10$^7$      \\
            9                                      & HCN            & J = 1-0                       & 88.631602          & 4.25             & 1.0 $\times$ 10$^6$      \\
            10                                     & HCN            & J = 3-2                       & 265.886434         & 25.52            & 6.8 $\times$ 10$^7$      \\
            11                                     & CS             & J = 2-1                       & 97.980953          & 7.05             & 3.3 $\times$ 10$^7$      \\
            12                                     & H$_2$CO        & J$_{kk'}$ = 2$_{12}$-1$_{11}$ & 140.839518         & 21.9             & 7.7 $\times$ 10$^5$      \\
            13                                     & CN             & N = 2-1,J=3/2-1/2,F=5/2-3/2   & 226.659558         & 16.31            & 1.3 $\times$ 10$^7$      \\
            14                                     & C$^{18}$O      & J = 2-1                       & 219.560358         & 15.81            & 9.3 $\times$ 10$^3$      \\
            15                                     & $^{13}$CO      & J = 3-2                       & 330.587965         & 31.73            & 3.3 $\times$ 10$^4$      \\
            16                                     & CH$_3$CN       & J=19-18 K=3                   & 349.393297         & 232.0            & 1.6 $\times$ 10$^7$      \\
            \hline\noalign{\smallskip}
            \multicolumn{6}{c}{optically thin lines}                                                                                                                   \\
            \hline
            1                                      & N$_2$H$^+$     & J = 1-0                       & 93.173404          & 4.47             & 1.4 $\times$ 10$^5$      \\
            2                                      & H$^{13}$CO$^+$ & J = 1-0                       & 86.754288          & 4.16             & 1.5 $\times$ 10$^5$      \\
            3                                      & H$^{13}$CO$^+$ & J = 3-2                       & 260.255339         & 24.98            & 3.2 $\times$ 10$^6$      \\
            4                                      & C$^{18}$O      & J = 1-0                       & 109.782176         & 5.27             & 1.9 $\times$ 10$^3$      \\
            5                                      & C$^{18}$O      & J = 2-1                       & 219.560358         & 15.81            & 9.3 $\times$ 10$^3$      \\
            6                                      & C$^{17}$O      & J = 3-2                       & 337.061130         & 32.35            & 3.5 $\times$ 10$^4$      \\
            \hline
         \end{tabular}
      }
   \end{center}
   \tablecomments{0.9\textwidth}{\textit{a}: The rest frequency of the line and upper level of energy values are obtained from the splatalogue database. \textit{b}: the critical density are derived from the formula $ n_{\text {crit }}=A_{u l} / \gamma_{u l} $.}
\end{table}

% %               one-column-spanning table
\begin{table}[htbp]
   \begin{center}
      \caption[]{The Number of Sources Observed by Tracers}\label{Tab:sources-lines}
      %%Please Capitalize the First Letter of Each Notional Word in table's caption
      \resizebox{1\textwidth}{!}{   % 自动调整表格宽度 
         \begin{tabular}{ccccccc}
            \hline\noalign{\smallskip}
            Serial Number & Spectral Line                   & Number of Sources & Reference                                  \\
            \hline\noalign{\smallskip}
            \multicolumn{4}{c}{optically thick lines}                                                                        \\
            \hline
            1             & HCO$^+$(1-0)                    & 350               & 3,5,10,12,13,14,15,16,17,18,19,20,21,27,28 \\
            2             & HCO$^+$(3-2)                    & 30                & 2,6,7,9,19,20,22,23,24, 25,27,30,31        \\
            3             & HCO$^+$(4-3)                    & 18                & 13,22,23,27,30                             \\
            4             & CO(2-1)                         & 24                & 13                                         \\
            5             & CO(3-2)                         & 8                 & 13                                         \\
            6             & CO(4-3)                         & 18                & 1,29                                       \\
            7             & HNC(1-0)                        & 129               & 12,28                                      \\
            8             & HNC(4-3)                        & 1                 & 22,23,30                                   \\
            9             & HCN(1-0)                        & 48                & 3,26,31                                    \\
            10            & HCN(3-2)                        & 8                 & 5,9,11,14,15,19,20                         \\
            11            & CS(2-1)                         & 24                & 6,19,20,21,22,23,24,30,31                  \\
            12            & H$_2$CO(2$_{12}$-1$_{11}$)      & 26                & 22,23,27,30                                \\
            13            & CN(N = 2-1,J=3/2-1/2,F=5/2-3/2) & 3                 & 5,8,10,14,16,20                            \\
            14            & C$^{18}$O(2-1)                  & 1                 & 7                                          \\
            15            & $^{13}$CO(3-2)                  & 1                 & 10,14,16,20                                \\
            16            & CH$_3$CN(J=19-18,K=3)           & 1                 & 4                                          \\
            \hline\noalign{\smallskip}
            \multicolumn{4}{c}{optically thin lines}                                                                         \\
            \hline
            1             & N$_2$H$^+$(1-0)                 & 304               & 6,12,22,26,27,28                           \\
            2             & H$^{13}$CO$^+$(1-0)             & 56                & 13,14                                      \\
            3             & H$^{13}$CO$^+$(3-2)             & 32                & 19,25,27                                   \\
            4             & C$^{18}$O(1-0)                  & 77                & 3,6,20                                     \\
            5             & C$^{18}$O(2-1)                  & 33                & 13                                         \\
            6             & C$^{17}$O(3-2)                  & 18                & 1                                          \\
            \hline\noalign{\smallskip}
         \end{tabular}
      }
   \end{center}
   \tablecomments{0.95\textwidth}{references are: 1: \citet{Yue2021}, 2: \citet{Contreras2018}, 3: \citet{Yang2020b}, 4: \citet{Liu2020}, 5: \citet{Liu2013}, 6: \citet{Lee2001}, 7: \citet{Tang2019}, 8: \citet{Qiu2012}, 9: \citet{Qin2016}, 10: \citet{Wu2009}, 11: \citet{Liu2011}, 12: \citet{He2015,He2016}, 13: \citet{Rygl2013}, 14: \citet{Klaassen2012}, 15: \citet{Klaassen2007}, 16: \citet{Klaassen2008}, 17: \citet{Sridharan2002}, 18: \citet{Kurtz1994}, 19: \citet{Reiter2011}, 20: \citet{Sun2009}, 21: \citet{Shirley2003}, 22: \citet{Mardones1997}, 23: \citet{Su2019}, 24: \citet{Lee2011}, 25: \citet{Gregersen1997,Gregersen2000}, 26: \citet{Kim2021}, 27: \citet{Fuller2005}, 28: \citet{Saral2018}, 29: \citet{Faundez2004}, 30: \citet{DiFrancesco2001}, 31: \citet{Keown2016}.}
\end{table}

\subsection{The Catalog}
\label{subsec.the catalog}

Up to May 2021, a total of 456 infall candidates have been identified, including 352 (77\%) sources being associated with high-mass star-forming regions, and 55 (12\%) with low mass, the remaining 49 (11\%) being uncertain. The low-mass sources are fewer, probably due to the more attention has been paid to high-mass sources in recent years, meanwhile the weakness of infall characteristics of low-mass sources. Of these sources, eight are observed by interferometric array, accounting for a small proportion and will not be discussed separately, the others are observed by single-dish telescope.

We divide the sources into clumps and cores as their properties may be different in infall motion. \citet*{He2015} counted the sources and found that more than 96.6 per cent of clumps have masses higher than 100 $M_\odot$. \citet*{Saral2018} thought that the dense cores have sizes \textless 0.1 pc and masses \textless 100 $M_\odot$. This paper adopts such a grouping criterion: when the radius of a source lower than 0.1pc (main criteria) and mass lower than 100 $M_\odot$, we believe this source is a core, otherwise, it is classified as a clump. For those sources whose radius data are not given in the literature, we use the calculated distance and the spatial resolution of the telescope to estimate the radius. Of all the 456 sources, 396 (87\%) are clumps and 60 (13\%) are cores. For 352 high-mass sources, 343 (97\%) are clumps, and for 54 low-mass sources, 46 (85\%) are cores. That is, vast majority of the high-mass sources are clumps, and most of the low-mass sources are cores. Therefore, we ignore the influence of these few parts and only discuss the statistical properties of high-mass clumps and low-mass cores. The difference of statistical result is less than 5\% when the samples are roughly divided into high-mass and low-mass sources compared with which are grouped into clumps and cores.

Table~\ref{Tab:Catalog-brief} lists the first ten sources of the infall catalog and their physical parameters. The complete catalog is in the appendix~\ref{Tab:Catalog}. All the parameters are fetched from the literature, except the distance.
The columns are as following:\\
(1): internal serial number. Source that observed by interferometric array are signed with $^*$ in the upper right. The symbol $^\dagger$ in the upper left indicates that the source is considered as a core, and the rest are clumps; \\
(2)(3): the right ascension and declination (J2000); \\
(4): the alias names of sources based on Galactic Coordinate; \\
(5): the distance of each source, which are calculated by the parallax-based distance estimator\footnote{\url{http://bessel.vlbi-astrometry.org/node/378}} (\citealt{Reid2016,Reid2019}). It used the Bayesian approach to assign sources to spiral arms based on their (\textit{l, b, v}) coordinates with respect to arm signatures seen in CO and {H\small \hspace{0.1em}I} surveys. The estimator gives two most likely values of distance from the full distance probability density function, and use the parameter $P$ to control the prior probability of obtaining a near/far distance.
We also calculated the scale hight of each source by using the maximum likelihood distance. We found that for some sources, the scale hight will be much larger than that of molecular gas disk, that is, $\sim$280 pc (\citealt{Su2021}). In such cases the near distance is adopted, although the probability of the far distance given by the estimator is greater; \\
(6): the velocity corresponding to the peak position of the optically thin line. Some sources have been observed by more than one groups, they may give different parameters in such cases, we adopt the most recent one. This applies to columns 6 to 13, except column 10;\\
(7)(8): the column density and number density (assumed to be spherical) of H$_2$; \\
(9): mass;\\
(10): optically thick lines that show blue asymmetry. The number is the same as that in table~\ref{Tab:spectral-line}, and the bold font indicate the one being used to calculate  $\delta v$;\\
(11): measure of asymmetry of the spectral profile. If multiple optically thick lines are used to identify infall motion, and multiple values are obtained accordingly, then we choose the one considered to be most effective in the paper, see section~\ref{subsec:Line profile};\\
(12): infall velocity; \\
(13): mass infall rate; \\
(14): the mass type of the source, \textit{high} and \textit{low} denote high-mass and low-mass, respectively; \\
(15): evolutionary stage. The number 1, 2, 3 refer to stages of prestellar, protostellar and {H\small \hspace{0.1em}II} region for high-mass sources and 1, 2 for low-mass ones. Sources labelled as C{H\small \hspace{0.1em}II} (compact {H\small \hspace{0.1em}II}) or HC{H\small \hspace{0.1em}II} (hyper-compact {H\small \hspace{0.1em}II}) or UC{H\small \hspace{0.1em}II} (ultra-compact {H\small \hspace{0.1em}II}) in the literature are all merged into category 3 in this work; \\
(16): The maser associated with these sources in the range of $1'$. The number 1, 2, 3, 4 represent the maser of CH$_3$OH--I, CH$_3$OH--II, H$_2$O, OH, respectively;\\
(17): The sources associated with outflow provided in some literature (\citealt{Wu2004,Maud2015,Li2019a,Zhang2020}) within $2'$; \\
(18): references. The number is the same as listed in table~\ref{Tab:sources-lines}. The angular resolution and sensitivity of adopted observations are listed below according to the sequence of references. 1: $13''$, 0.19 K at 0.27 km s$^{-1}$ for CO (4-3) and 0.14 K at 0.27 km s$^{-1}$ for C$^{17}$O (3-2), 2: $1.2''$, 0.14 K at 0.08 km s$^{-1}$, 3: $62''$, 0.1 K at 0.2 km s$^{-1}$, 4: $0.14''$, 5: $18.3''$, 0.15 K for HCN (3-2), 6: $52''$, 7: $3.74'' \times 3.04''$ (-86$^\circ$) for C$^{18}$O (2-1), $0.86'' \times 0.86''$ (85$^\circ$) for HCO$^+$(3-2), 8: $1.4'' \times 1.3''$ (57$^\circ$) for CN (N=2-1), 9: $2.3'' \times 2.1''$ (-60$^\circ$), 11: $1.63'' \times 1.28''$ (-81.4$^\circ$), 12: $19.2''$, 13: $28''$, 0.10 K at 0.08 km s$^{-1}$ for HCO$^+$(1-0), and $11''$, 1.3 K at 1.5 km s$^{-1}$ for CO (2-1), 14: $15''$, 0.1 K at 0.42 km s$^{-1}$ (356 GHz) and 0.07 K at 0.82 km s$^{-1}$ (347 GHz), 15: $14''$, 0.13 K at 0.53 km s$^{-1}$, 16: $14''$, 0.13 K at 0.42 km s$^{-1}$, 19: $18''$, 20: $55''$, $\sim$0.07 K, 21: $24.5''$, 22: $28''$, 23: $16.55''$, 24: $52''$, 25: $26''$, 26: $32''$, $\sim$0.4 K at 0.11 km s$^{-1}$, 27: $29''$ (IRAM), $14''$ (JCMT), 0.08 K for HCO$^+$(4-3) and 0.07 K for HCO$^+$(1-0), 28: $38''$, 29: $24''$, 30: $\sim2''$, 31: $9.4''$, 0.02 K at 0.044 km s$^{-1}$.\\

\begin{table}[htbp]
   % \begin{center}
   \setlength\LTleft{-0.05in}               % 设置左边距，-1为超出左边距-1
   \setlength\LTright{2in plus 1 fill}     % 设置右边距 (好像不起作用)
   \setlength{\tabcolsep}{4pt}           % 设置列间距,默认为6pt,可以利用列间距来间接控制右边距
   \caption[]{The Catalog of All Collected Infall Sources}\label{Tab:Catalog-brief}
   %%Please Capitalize the First Letter of Each Notional Word in table's caption
   \resizebox{1.05\textwidth}{!}{   % 自动调整表格宽度     
      \begin{tabular}{cccccccccccccccccl}
         \hline\noalign{\smallskip}
         \hline\noalign{\smallskip}

         No.              & R.A.                 & Dec.                 & Alias             & Dist.              & $V_{\text{LSR}}$           & $N$(H$_2$)               & $n$(H$_2$)               & Mass                     & Lines             & $\delta v$        & $V_{in}$                   & $\dot M_{in}$                      & Type              & Stage             & Maser             & Outflow           & Ref               \\
                          & \scriptsize{(J2000)} & \scriptsize{(J2000)} &                   & \scriptsize{(kpc)} & \scriptsize{(km s$^{-1}$)} & \scriptsize{(cm$^{-2}$)} & \scriptsize{(cm$^{-3}$)} & \scriptsize{(M$_\odot$)} &                   &                   & \scriptsize{(km s$^{-1}$)} & \scriptsize{(M$_\odot$ yr$^{-1}$)} &                   &                   &                   &                   &                   \\
         \scriptsize{(1)} & \scriptsize{(2)}     & \scriptsize{(3)}     & \scriptsize{(4)}  & \scriptsize{(5)}   & \scriptsize{(6)}           & \scriptsize{(7)}         & \scriptsize{(8)}         & \scriptsize{(9)}         & \scriptsize{(10)} & \scriptsize{(11)} & \scriptsize{(12)}          & \scriptsize{(13)}                  & \scriptsize{(14)} & \scriptsize{(15)} & \scriptsize{(16)} & \scriptsize{(17)} & \scriptsize{(18)} \\

         \hline
         1                & 00:35:40.1           & +66:14:23            & G121.34+3.42      & 0.18               & -5.3                       & 9.8E+21                  &                          &                          & \textbf{1},9      & -0.47             &                            &                                    &                   &                   &                   &                   & 3                 \\
         2                & 00:36:47.5           & +63:29:01            & G121.30+0.66      & 0.9                & -17.6                      &                          &                          &                          & 2                 & -0.54             &                            &                                    & high              &                   & 123               & Y                 & 19                \\
         3                & 00:36:53.6           & +63:28:03            & G121.31+0.64      & 1.1                & -17.3                      & 1.7E+22                  &                          &                          & \textbf{1},9      & -0.63             & 0.29                       & 1.6E-3                             &                   & 2                 &                   &                   & 3                 \\
         4                & 00:52:25.1           & +56:33:54            & G123.07-6.31      & 2.2                & -30.4                      &                          &                          &                          & 1,\textbf{2},10   & -0.20             &                            &                                    & high              & 3                 & 123               & Y                 & 19,20             \\
         5                & 02:19:51.8           & +61:03:26            & G133.42+0.00      & 0.9                & -15.2                      & 4.8E+21                  &                          &                          & 9                 & -0.59             &                            &                                    &                   & 2                 &                   &                   & 3                 \\
         6                & 02:27:03.8           & +61:52:25            & W3(OH)            & 2.4                & -48.0                      &                          & 6.0E+6                   &                          & \textbf{1},10     & -0.23             & 0.06                       & 3.0E-4                             & high              & 3                 & 1234              & Y                 & 14,15,19,20       \\
         $^\dagger$7$^*$  & 02:27:04.7           & +61:52:26            & W3(H$_2$O)        & 2.0                & -47.8                      & 2.6E+24                  & 1.0E+8                   & 2.6E+1                   & 2,\textbf{10}     & -0.65             & 2.70                       & 2.3E-3                             & high              & 2                 & 1234              & Y                 & 9                 \\
         $^\dagger$8      & 02:53:12.2           & +68:55:52            & LDN 1355          & 0.3                & -3.8                       & 1.4E+21                  & 9.1E+3                   &                          & 11                & -0.42             & 0.12                       & 8.4E-6                             & low               & 1                 &                   &                   & 6,24              \\
         $^\dagger$9      & 03:26:37.0           & +30:15:26            & IRAS03235+3004    & 0.35               & 4.9                        &                          &                          &                          & 2                 & -0.66             &                            &                                    & low               & 2                 & 3                 &                   & 25                \\
         $^\dagger$10     & 03:28:32.5           & +31:11:05            & J03283258+3111040 & 0.3                & 7.2                        &                          &                          &                          & 9                 & -0.45             & 0.17                       & 6.5E-6                             & low               & 2                 &                   &                   & 26                \\

         \hline\noalign{\smallskip}
      \end{tabular}
   }
   % \end{center}
\end{table}

\section{Statistics and Distribution}           %% first-level sections will be auto-capitalized
\label{section:Basic Properties}

\subsection{Spatial Distribution}

\begin{figure}[htbp]
   \centering
   \includegraphics[width=10cm, angle=0]{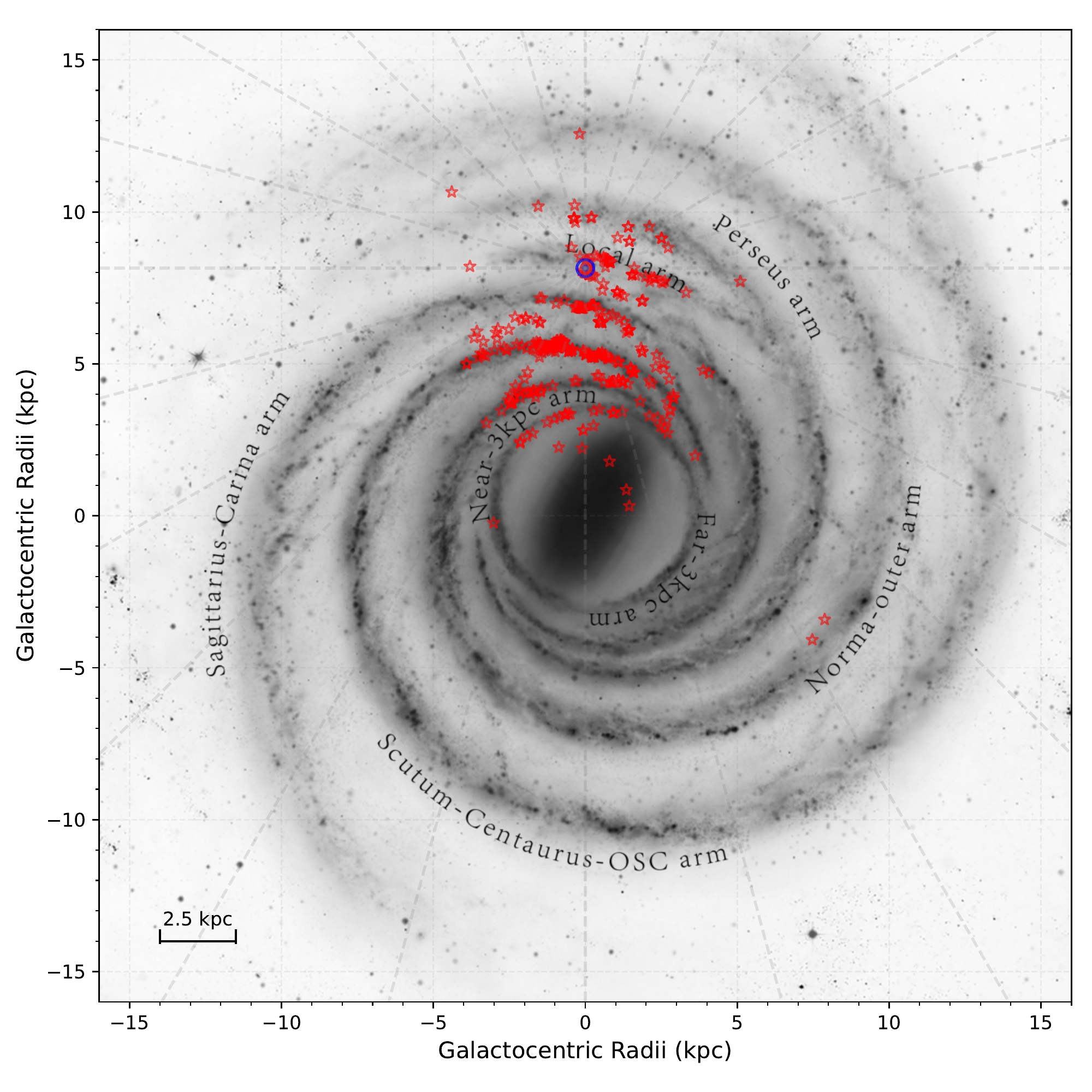}
   % \begin{minipage}[]{85mm}
   \caption{\small Galactic distribution of the infall sources. The blue circle indicates the location of the Sun. The red pentagram represents the location of the infall sources. The background is the imaginary face-on image of the Milky Way (created: ``Xing-Wu Zheng \& Mark Reid  BeSSeL/NJU/CFA'').}
   %\end{minipage}
   \label{fig:Galactic-distribution}
\end{figure}

Figure~\ref{fig:Galactic-distribution} shows the galactic distribution of all infall sources overlaid on the imaginary face-on image of the Milky Way, which is adopted from ``Xing-Wu Zheng \& Mark Reid  BeSSeL/NJU/CFA\footnote{\url{https://astronomy.nju.edu.cn/xtzl/EN/}}'', and which is believed to be currently the most scientifically accurate visualization of the Milky Way. The blue circle indicates the location of the Sun, with a distance 8.15 kpc to the Galactic center (\citealt{Reid2019}). The red pentagram represents the location of the infall sources. The scale of the image is indicated at lower left corner of the figure.
We found that most of the sources (about 90\%) are located within 5 kpc of the Sun. This may be caused by the observation effect. Moreover, The numbers of infall sources inside and outside the solar circle are 390 and 66, respectively.
Although the observations to date are far from complete, the present evidence shows that the star formation activity inside the solar circle is more intense than the outside (\citealt{Djordjevic2019}), which needs further investigation in the future.
Some infall sources are distributed on the spurs between the Local and Sagittarius arms (\citealt{Xu2016}) and between the Sagittarius and Scutum arms.

\subsection{Excitation Temperature}
\label{Tex} %  T$_{ex}$

The excitation temperature $T_{ex}$ is usually estimated using the standard radiation transfer equation and assuming local thermodynamic equilibrium (LTE).
% T$_{ex}$ can be calculated as:
% \begin{equation}
%    T_{\mathrm{ex}}=\frac{h v / k}{\ln \left[\frac{h v / k}{\left(A \tau_{\mathrm{m}} / \tau\right)+J_{\nu}\left(T_{\mathrm{bg}}\right)}+1\right]}
% \end{equation}
% where $h$ is the Planck constant, $k$ is the Boltzmann constant, $\nu$ is the frequency of the observed transition, and $J\nu(T_{bg})$ is the Planck radiation which was calculated by the following equation
% \begin{equation}
%    J_{v}\left(T_{\mathrm{bg}}\right)=\frac{h v / k}{e^{h v / k T_{\mathrm{bg}}}-1}
% \end{equation}
% Where the cosmic background radiation temperature $T_{bg}$ is assumed to be 2.726 K.

\begin{figure}[htbp]  %[htbp]中的h是浮动的意思，H表示固定
   \centering    %居中

   \subfloat[Distribution of all $T_{ex}$] %第一张子图，[]内可填写子图的标注
   {
      \begin{minipage}[t]{0.5\linewidth} %子图以0.33倍行宽显示
         \centering          %子图居中
         \includegraphics[width=1\linewidth]{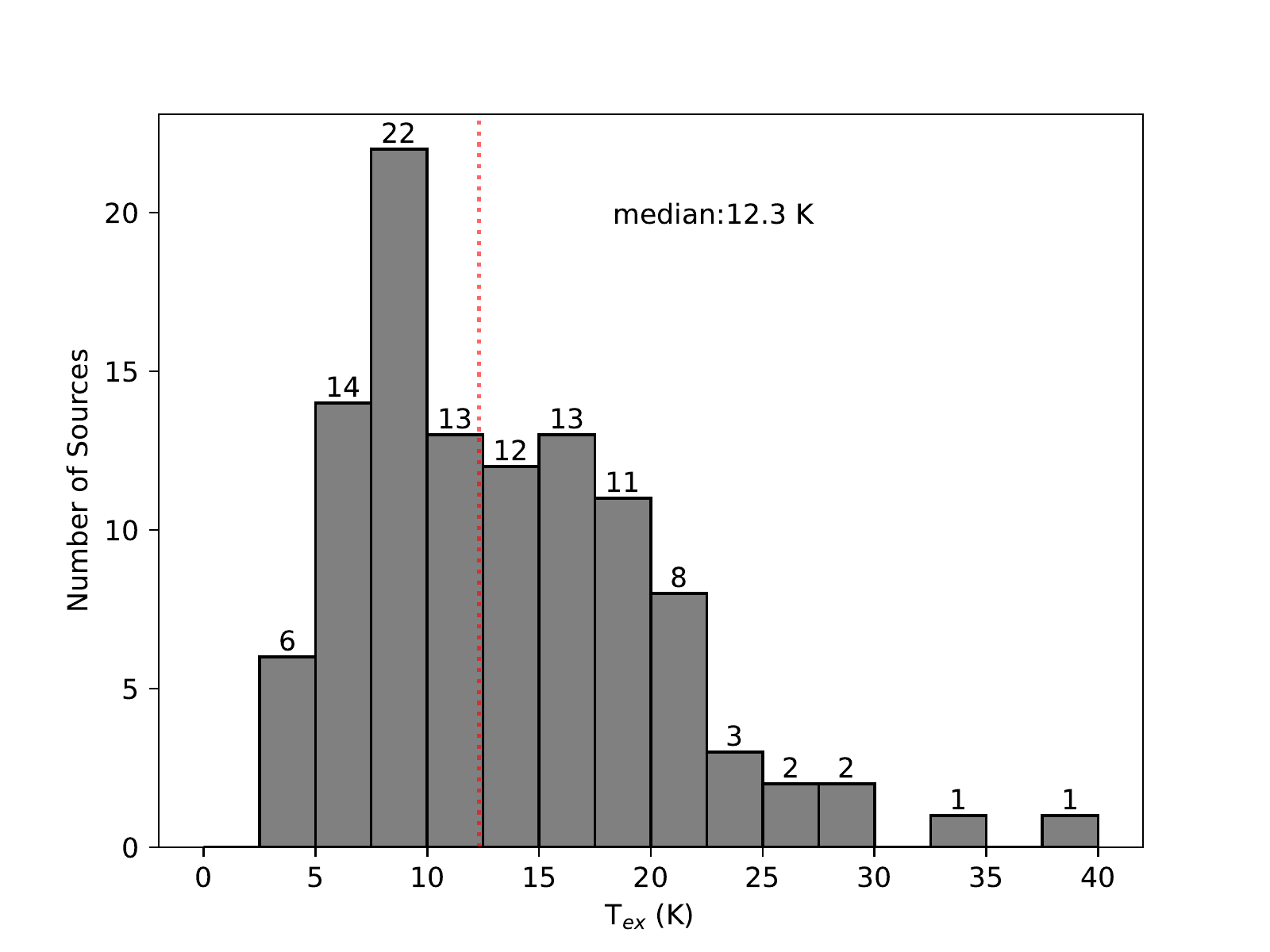}   %子图空间内以行宽的0.5倍大小显示
         \captionsetup{font={scriptsize}}
      \end{minipage}
   } % 注意这里不能回车空行，否则两张图会上下排列，而不是并排排列
   \subfloat[Distribution of $T_{ex}$ of each stage] %第二张子图
   {
      \begin{minipage}[t]{0.5\linewidth} %子图以0.33倍行宽显示
         \centering      %子图居中
         \includegraphics[width=1\linewidth]{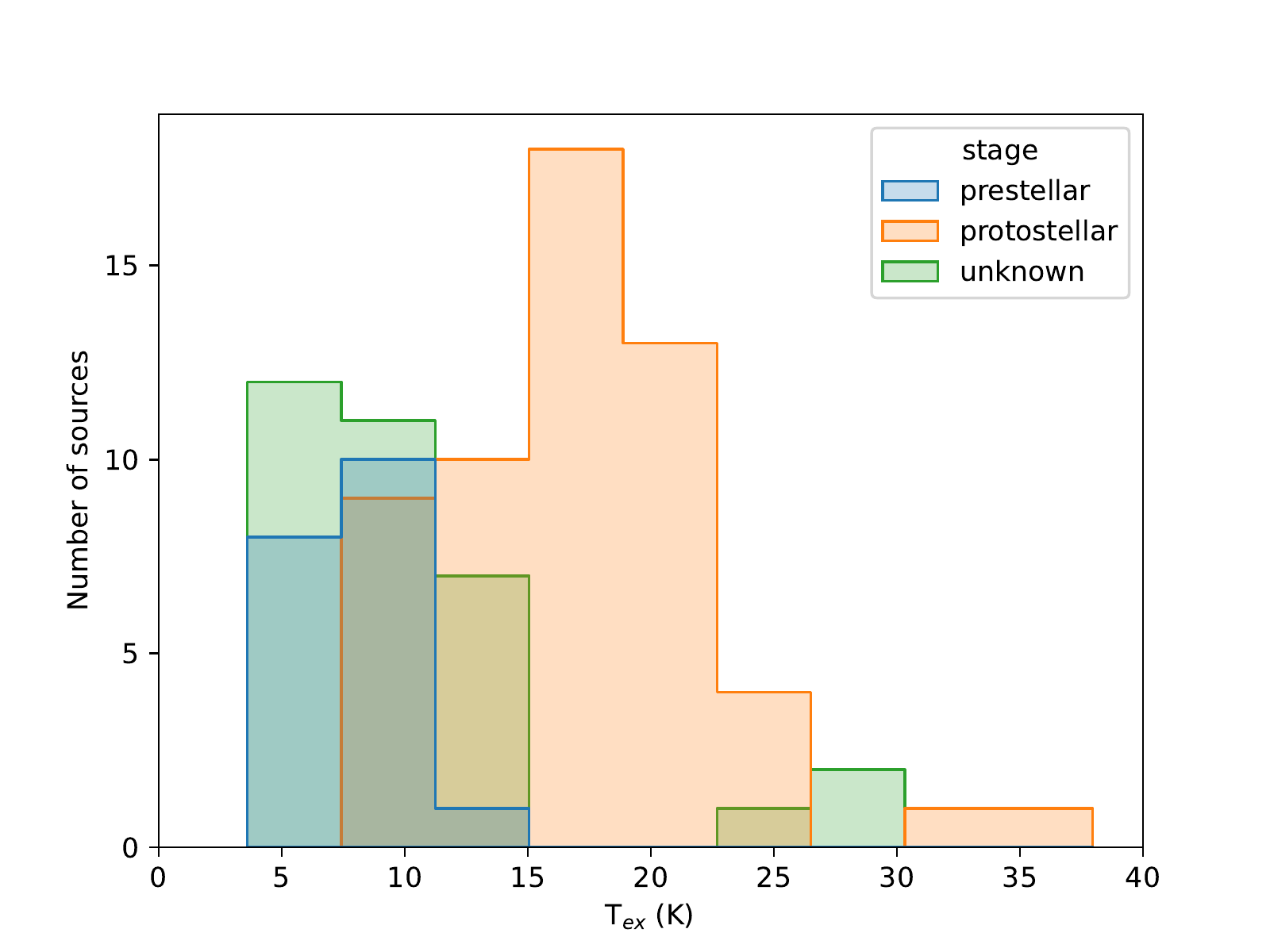}   %子图空间内以行宽的0.5倍大小显示
         \captionsetup{font={scriptsize}}
      \end{minipage}
   }%
   \caption{\small Distribution of $T_{ex}$}  %整个大图的标注
   \captionsetup{font={scriptsize}}
   \label{fig:Tex}  %图片引用标记
\end{figure}

Of all the sources, there are 109 that have valid $T_{ex}$ values, all of them are fetched from the literature. we don't differentiate these sources by high mass or low mass because of the relatively small number. The median and mean value of $T_{ex}$ are 12.3 K and 13.4 K respectively. Figure~\ref{fig:Tex} (a) shows the distribution of $T_{ex}$ of all, and (b) shows the distribution of $T_{ex}$ of each stage. It can be seen from the figure~\ref{fig:Tex} (b) that the excitation temperatures in the protostar stage are significantly higher than those in the prestellar stage, and their median values are 17.1 K and 7.8 K respectively. It may because of the fact that in the prestellar stage, it is too early to release enough gravitational energy to heat the surrounding environment. However, this trend should be interpreted with caution due to the small number of samples.

\begin{figure}[htbp]  %[htbp]中的h是浮动的意思，H表示固定
   \centering    %居中

   \subfloat[Distribution of all $N$(H$_2$)] %第一张子图，[]内可填写子图的标注
   {
      \begin{minipage}[t]{0.555\linewidth} %子图以0.555倍行宽显示
         \centering          %子图居中
         \includegraphics[width=0.95\linewidth]{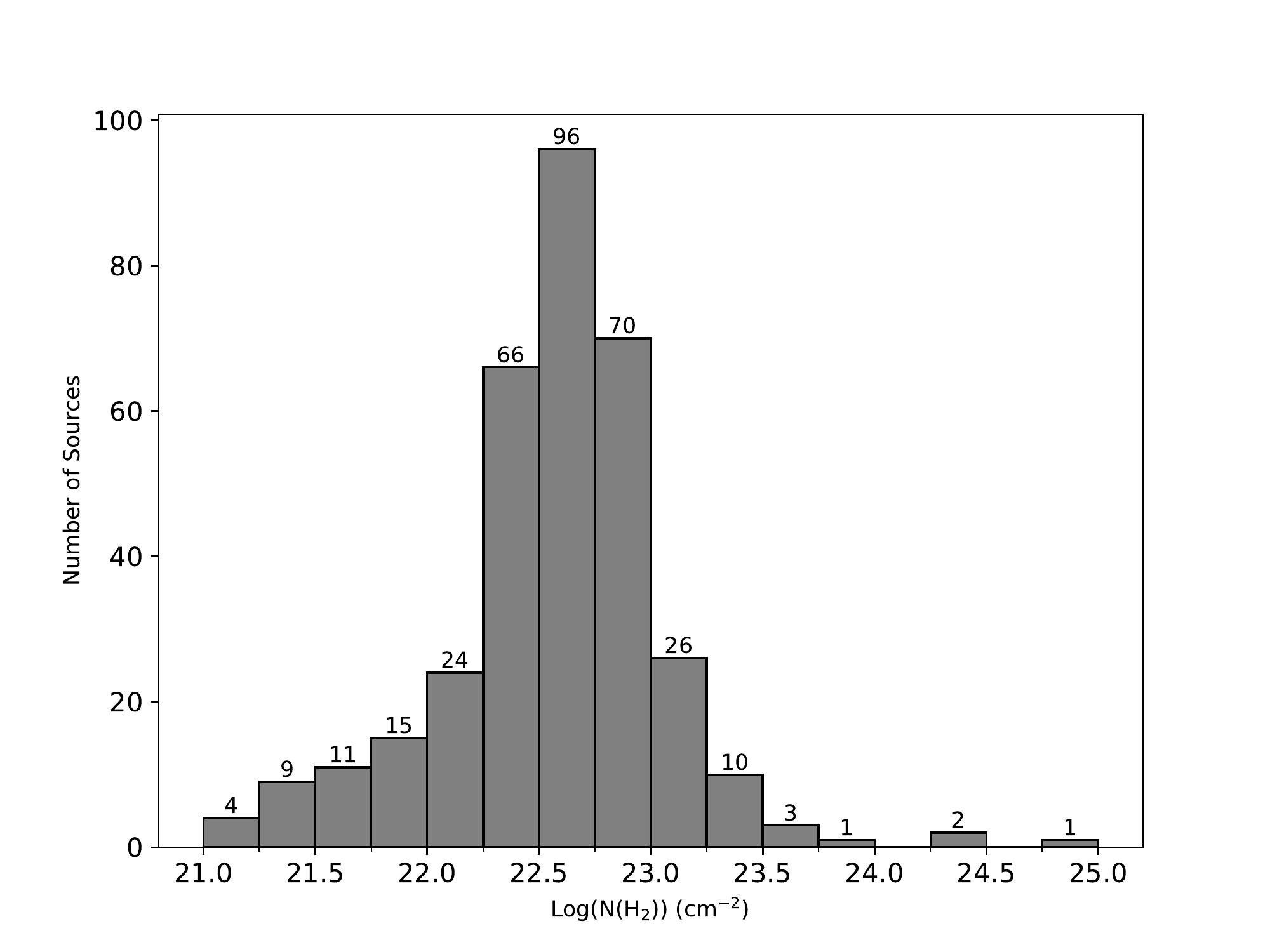}   %子图空间内以行宽的0.5*0.95倍大小显示
         \captionsetup{font={scriptsize}}
      \end{minipage}
   } % 注意这里不能回车空行，否则两张图会上下排列，而不是并排排列
   \subfloat[Distribution of high-mass clumps and low-mass \\ \hspace*{2em} cores of $N$(H$_2$)] %第二张子图
   {
      \begin{minipage}[t]{0.435\linewidth} %子图以0.435倍行宽显示
         \centering      %子图居中
         \includegraphics[width=1\linewidth]{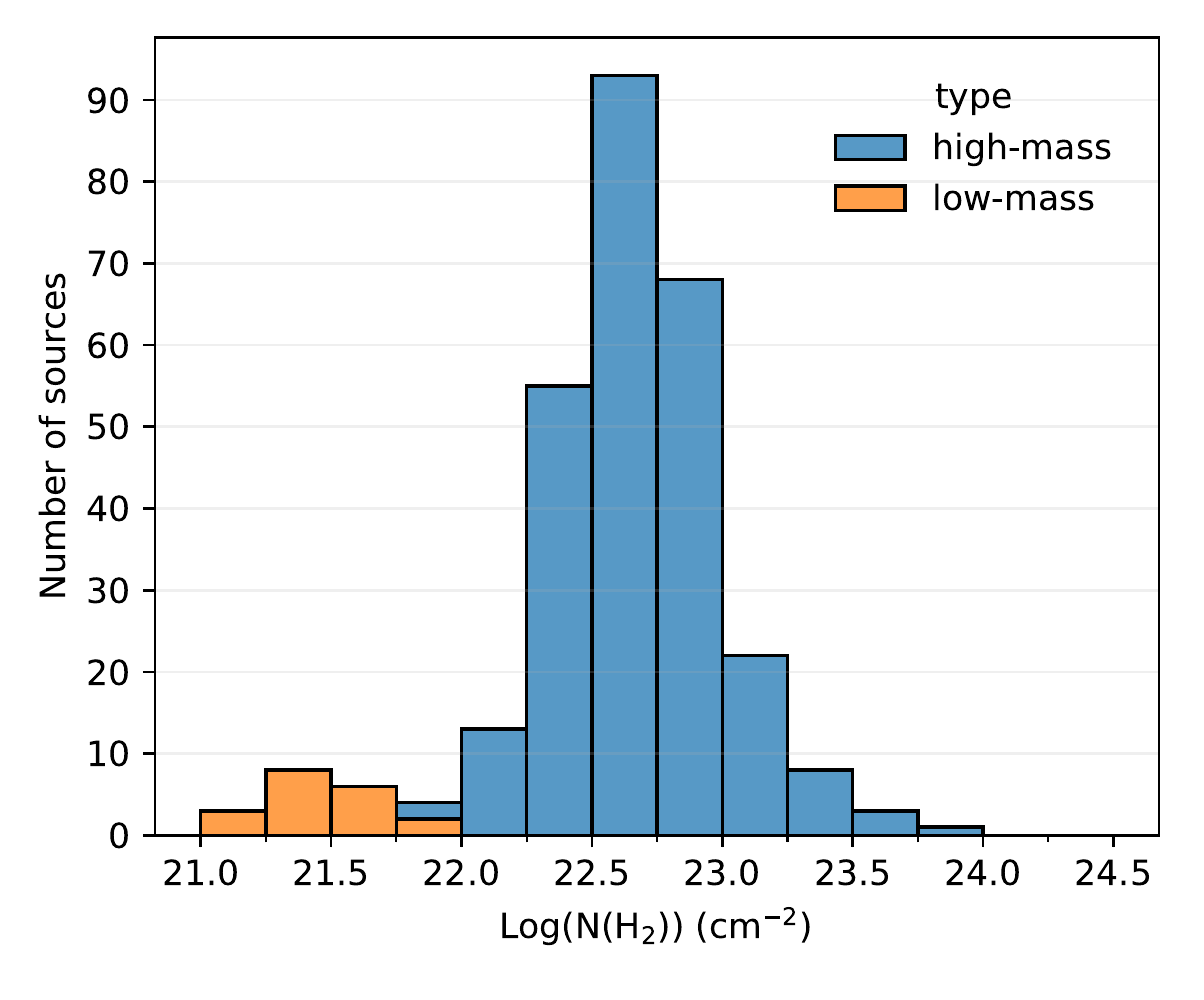}   %子图空间内以行宽的0.5倍大小显示
         \captionsetup{font={scriptsize}}
      \end{minipage}
   }% 
   \caption{\small Distribution of H$_2$ column density}  %整个大图的标注
   % \captionsetup{font={scriptsize}}
   \label{fig:column density}  %图片引用标记
\end{figure}

\begin{figure}[h]
   \centering
   \includegraphics[width=15cm, angle=0]{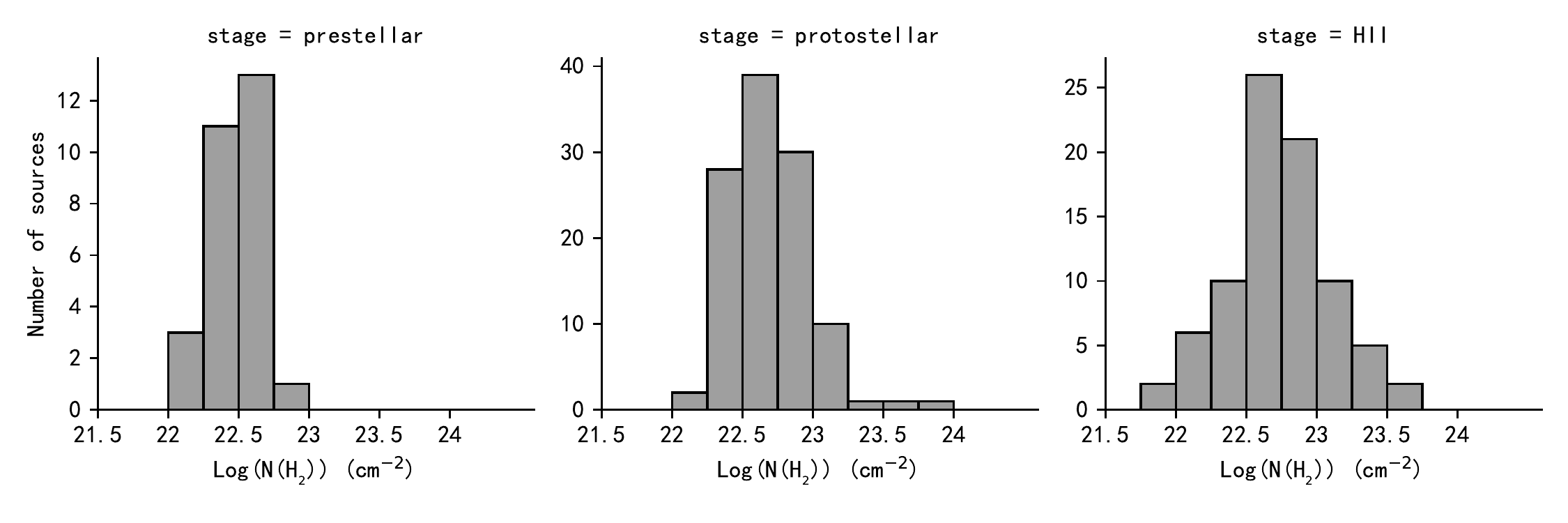}
   % \begin{minipage}[]{85mm}
   \caption{\small Distribution of H$_2$ column density for high-mass clumps at different evolutionary stages.}
   %\end{minipage}
   \label{fig:log_N_H2_stage}
\end{figure}

\subsection{Column Density}
\label{column density}

Figure~\ref{fig:column density} (a) shows the H$_2$ column density $N_{\text{H}_2}$ distribution of all sources with valid values given in the literature. The column density of H$_2$ ranges from 1.2$\times$ 10$^{21}$ to 9.8 $\times$ 10$^{24}$ cm$^{-2}$, with a median value of 4.17$\times$ 10$^{22}$ cm$^{-2}$.
Figure~\ref{fig:column density} (a) shows that $N_{\text{H}_2}$ is not log-normally distributed.
% This could be the following reasons: different methods for calculating column density are used in different papers; the results of column density calculated using different molecular lines are different based on abundance ratios to hydrogen molecules. 
Figure~\ref{fig:column density} (b) gives the column density distributions of low-mass and high-mass sources. It can be seen that the $N_{\text{H}_2}$ distribution of low-mass cores and high-mass clumps is significantly separated. Their median values are 2.6 $\times$ 10$^{21}$ and 4.8 $\times$ 10$^{22}$ cm$^{-2}$, respectively.
Among them, there are few samples of low-mass cores, so its distribution has little significance. For high-mass clumps, the distribution has a threshold of 10$^{22}$ $\sim$ 10$^{24}$ cm$^{-2}$, which is consistent with the literature (\citealt{Rygl2013,He2016}).
% The order of the $N_{\text{H}_2}$ for low-mass stars is 10$^{21}$ cm$^{-2}$, and for high-mass stars is about 10$^{22}$ cm$^{-2}$ $\sim$ 10$^{24}$ cm$^{-2}$.
Figure~\ref{fig:log_N_H2_stage} presents the distribution of $N_{\text{H}_2}$ for high-mass clumps at different evolutionary stages. The median values for high-mass are 3.1 $\times$ 10$^{22}$, 4.7 $\times$ 10$^{22}$, 5.4 $\times$ 10$^{22}$ cm$^{-2}$ for prestellar, protostellar, {H\small \hspace{0.1em}II}, respectively. Taking into account that the corresponding RMS are 0.18, 0.28, 0.35 in the case of logarithm, respectively, the column density of H$_2$ increases slightly with the evolutionary stage.
% Only the {H\small \hspace{0.1em}II} stage is similar to normal distribution.
Different methods for calculating column density are used in different papers, the results of column density calculated using different molecular lines are different based on abundance ratios to hydrogen molecules.
% Compared with \citet{He2016}, the corresponding median values are 3.09 $\times$ 10$^{22}$, 4.68$\times$ 10$^{22}$, 5.50$\times$ 10$^{22}$, 3.63 $\times$ 10$^{22}$.

\subsection{The Skewness Parameter }
\label{subsec:Line profile}

As stated in the introduction, the infall motion is signatured by the spectral line asymmetry of optically thick line and a single-peaked profile  (\citealt{Leung1977,Zhou1992,Mardones1997,Lee1999,Evans1999}).
To demonstrate the measure of asymmetry, \citet{Mardones1997} defined a non-dimensional parameter $\delta V$
\begin{equation}
   \delta V=\frac{V_{\text {thick }}-V_{\text {thin }}} { \Delta V_{\text {thin }}}
\end{equation}
where $V_{\text{thick}}$ and $V_{\text{thin}}$ are the velocities of the peaks of optically thick and thin lines. $\Delta V_{\text{thin}}$ is the line width (FWHM) of the optically thin line. The values $\delta V$ obtained from different line pairs may be different. When $\delta V \leqslant$ \ -0.25 or $\delta V \geqslant$ \ 0.25, it can usually be considered that the source has a significant blue or red profile.

Different molecules may give different result in studying infall.
\citet{Mardones1997} observed 47 protostars to identify infall sources with H$_2$CO (2$_{12}$ -- 1$_{11}$) and CS (2--1), and found that the fraction of blue profiles decrease from Class 0 to Class I.
However \citet{Gregersen1998} observed a larger sample which includes the Class I sources of \citet{Mardones1997}, but did not find such trend using HCO$^+$(3--2).
\citet{Fuller2005} observed HCO$^+$ J = 1 -- 0, J = 3 -- 2, J = 4 -- 3, and H$_2$CO (2$_{12}$ -- 1$_{11}$) toward 77 850 $\mu$m continuum sources, and  suggests HCO$^+$(1--0) and H$_2$CO (2$_{12}$ - 1$_{11}$) trace infall more effectively than HCO$^+$(3--2) and HCO$^+$(4--3).
\citet{Rygl2013} used HCO$^+$(1--0), HCO$^+$(4--3) and CO (3--2) to observe another sample of high infrared extinction clouds, came to a similar conclusion.
Therefore, the $\delta V $ values derived from HCO$^+$(1-0) and H$_2$CO (2$_{12}$ -- 1$_{11}$) are given with high priority.

\begin{figure}[h]
   \begin{minipage}[t]{0.4\linewidth}
      \centering
      \includegraphics[width=60mm]{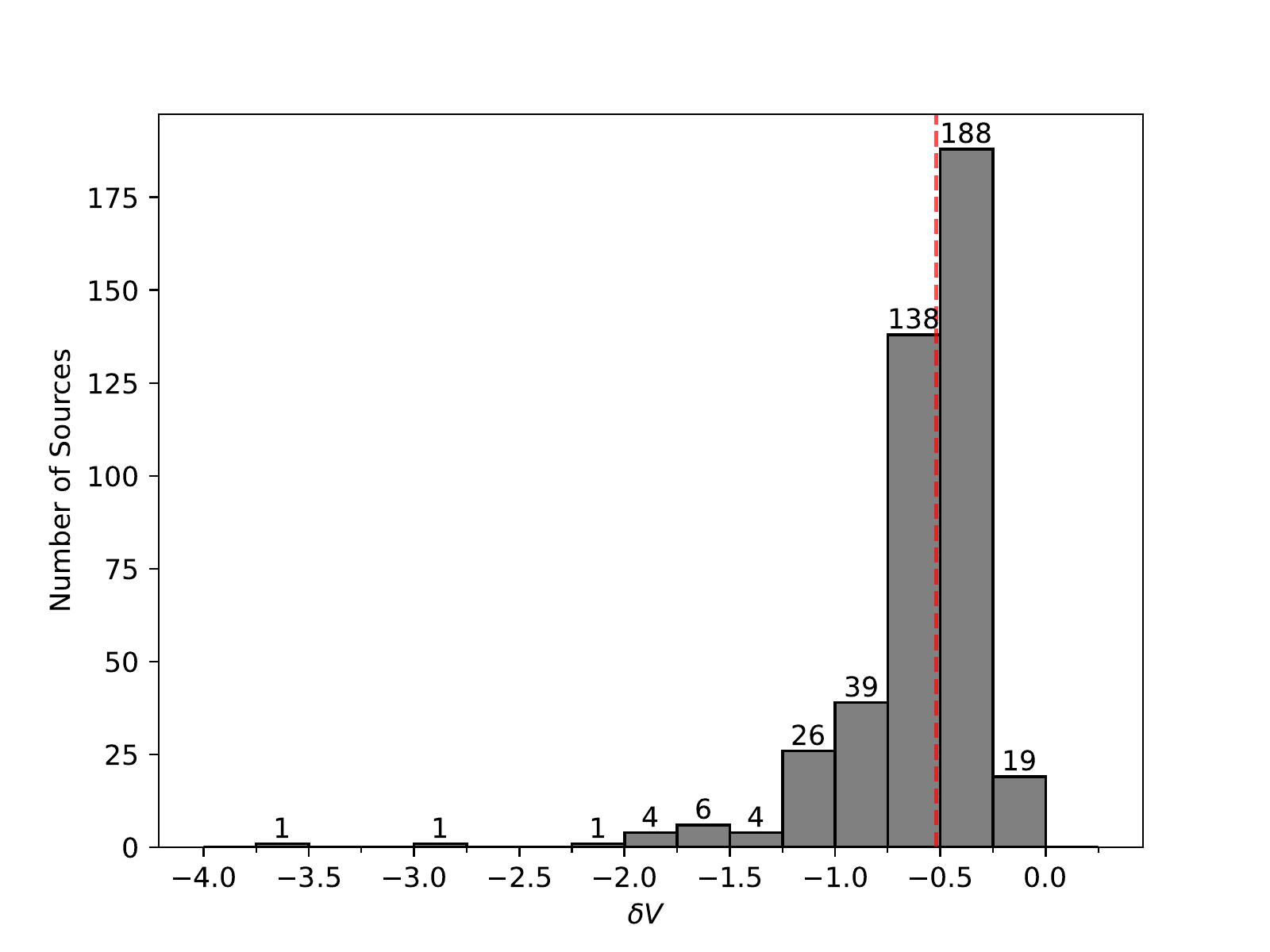}
      \caption{\small Distribution of $\delta V$ for all sources \\ given in the literature. The red dotted line \\represents the median value of $\delta V$.}
      \label{fig:deltaV}
   \end{minipage}%
   \begin{minipage}[t]{0.6\linewidth}
      \centering
      \includegraphics[width=80mm]{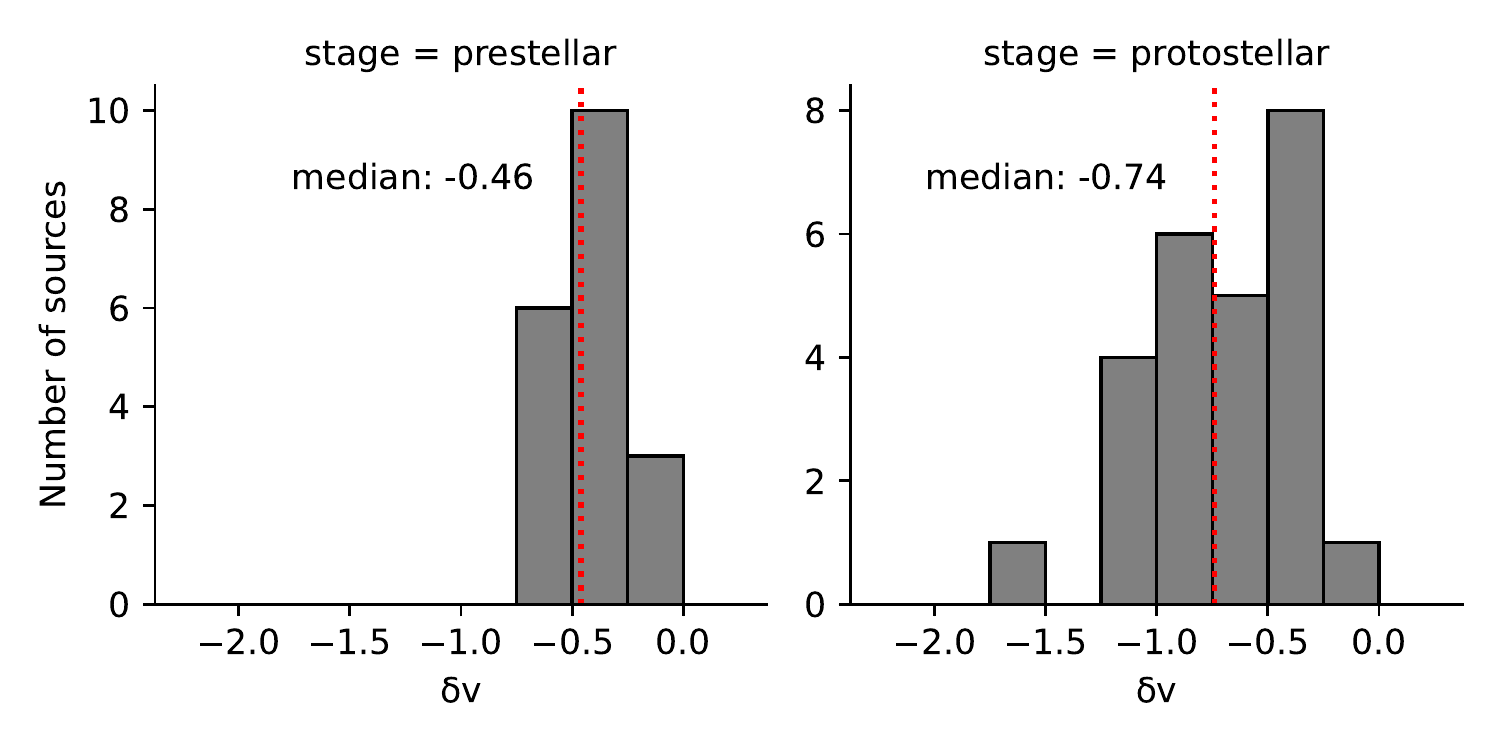}
      \caption{\small Distribution of $\delta V$ for low-mass cores. The red dotted line represents the median value of $\delta V$ for each stage.}
      \label{fig:deltaV-low}
   \end{minipage}%
\end{figure}

\begin{figure}[h]
   \centering
   \includegraphics[width=15cm, angle=0]{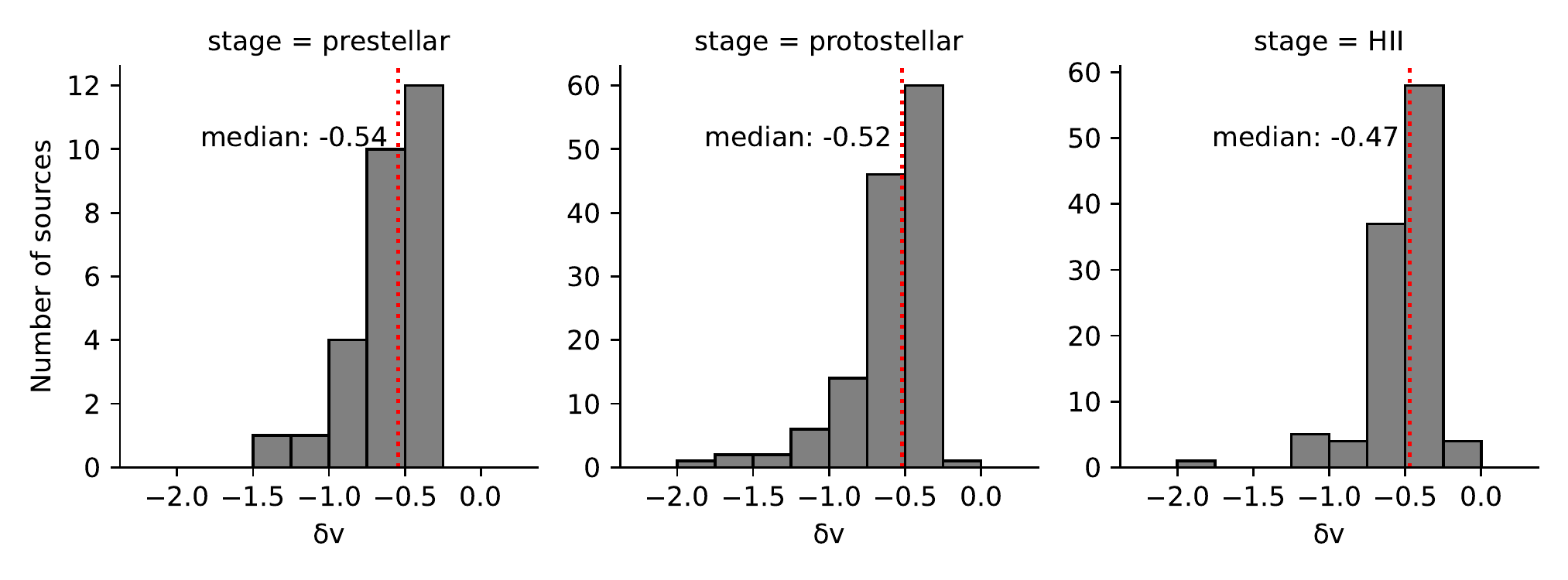}
   % \begin{minipage}[]{85mm}
   \caption{\small Distribution of $\delta V$ for high-mass clumps. The red dotted line represents the median value of $\delta V$ for each stage.}
   %\end{minipage}
   \label{fig:deltaV-high}
\end{figure}

Figure~\ref{fig:deltaV} shows the distribution of 429 valid $\delta V$ values of all sources. Although $\delta V$s are calculated from different lines, we believe this distribution has its scientific significance since the large proportion is identified with HCO$^+$(1-0).
The median value of $\delta V$ for all sources is -0.52. For high-mass clumps, the median value identified with HCO$^+$(1-0) is -0.50. In contrast, \citet{Rygl2013} derived -0.64 with the same line.
For low-mass cores, figure~\ref{fig:deltaV-low} presents the $\delta V$ distribution of different evolutionary stages. The median values at stages of prestellar and protostellar are -0.46 and -0.74, respectively. Considering the corresponding RMS are 0.16 and 0.33 respectively, the blue asymmetry may have no difference with evolution of low-mass cores.
Further observations are needed because of the small sample size.
As figure~\ref{fig:deltaV-high} shows, the $\delta V$ of high-mass infall clumps are -0.55, -0.52, -0.47 at the stage of prestellar, protostellar and {H\small \hspace{0.1em}II} region respectively. The RMS are 0.23, 0.29, 0.25, respectively. It seems that the blue asymmetry of high-mass infall clumps have the same trend as those of low-mass.

% 此方式是把多张图并排放在一起,因此图号会随着minipage的数量依次增加

\subsection{Infall Velocity}

So far, there are several different approaches to estimate the infall velocity: velocity difference, ``two-layer'' model, ``hill'' model, RATRAN, etc.
(i) Some works (e.g., \citealt{Liu2013,He2015,He2016,Su2019}) estimated roughly with $V_{\text{in}} = V_{\text{obs}} - V_{\text{sys}}$, where $V_{\text{obs}}$ means the velocity at blue peak of the optically thick line, $V_{\text{sys}}$ is the system velocity from Gaussian fit of the optically thin line.
(ii) The ``two-layer'' model (\citealt{Myers1996}) assumes that the excitation temperature in each layer is constant.
Some works (e.g., \citealp*{Lee2001,Qin2016,Tang2019,Yue2021}) used the ``two-layer'' model to calculate the infall velocity with the following equation when $V_{\text {in }} \ll \sigma (2 \ln \tau_0)^{1 / 2}$,
\begin{equation}
   V_{\text {in }}=\frac{\sigma^{2}}{V_{\text {red }}-V_{\text {blue }}} \ln \left(\frac{1+e^{\left(T_{\text {blue }}-T_{\text {dip }}\right) / T_{\text {dip }}}}{1+e^{\left(T_{\text {red }}-T_{\text {dip }}\right) / T_{\text {dip }}}}\right)
\end{equation}
where $V_{\text{blue}}$ and $V_{\text{red}}$ are the velocities of the blue and red peaks of optically thick line, $T_{\text{blue}}$ and $T_{\text{red}}$ are the corresponding peak intensities, $T_{\text{dip}}$ is the intensity of self-absorption dip between the two peaks, and $\sigma$ is the velocity dispersion of optically thin line.
(iii) The ``hill'' model assumes that the excitation temperature is increase inward as a linear function of optical depth (\citealt{DeVries2005}).
\citet{DeVries2005} thought that the ``two-layer'' model underestimates the infall velocity by a factor of $\sim$2 when matching the two-peak profiles while ``hill'' model matches better with an rms error of 0.01 km s$^{-1}$. Blue-asymmetric line profiles with shoulder or redshifted peak can not generally well fitted by the ``two-layer'' or ``hill'' model.
(iv) A more complex model is RATRAN, which uses 1D Monte Carlo radiative transfer code (\citealt{Hogerheijde2000}) to construct the infall model with more input parameters, such as the radius, mass, density profile, turbulent velocity, kinetic temperature distribution, and the abundance of the molecule. This model can better constrain infall velocity than ``two-layer'' model (\citealt{Peretto2013}).
The infall velocities derived by different approaches are relatively uniform and will not have a significant impact on the statistical results.

\begin{figure}[h]
   \begin{minipage}[t]{0.4\linewidth}
      \centering
      \includegraphics[width=60mm]{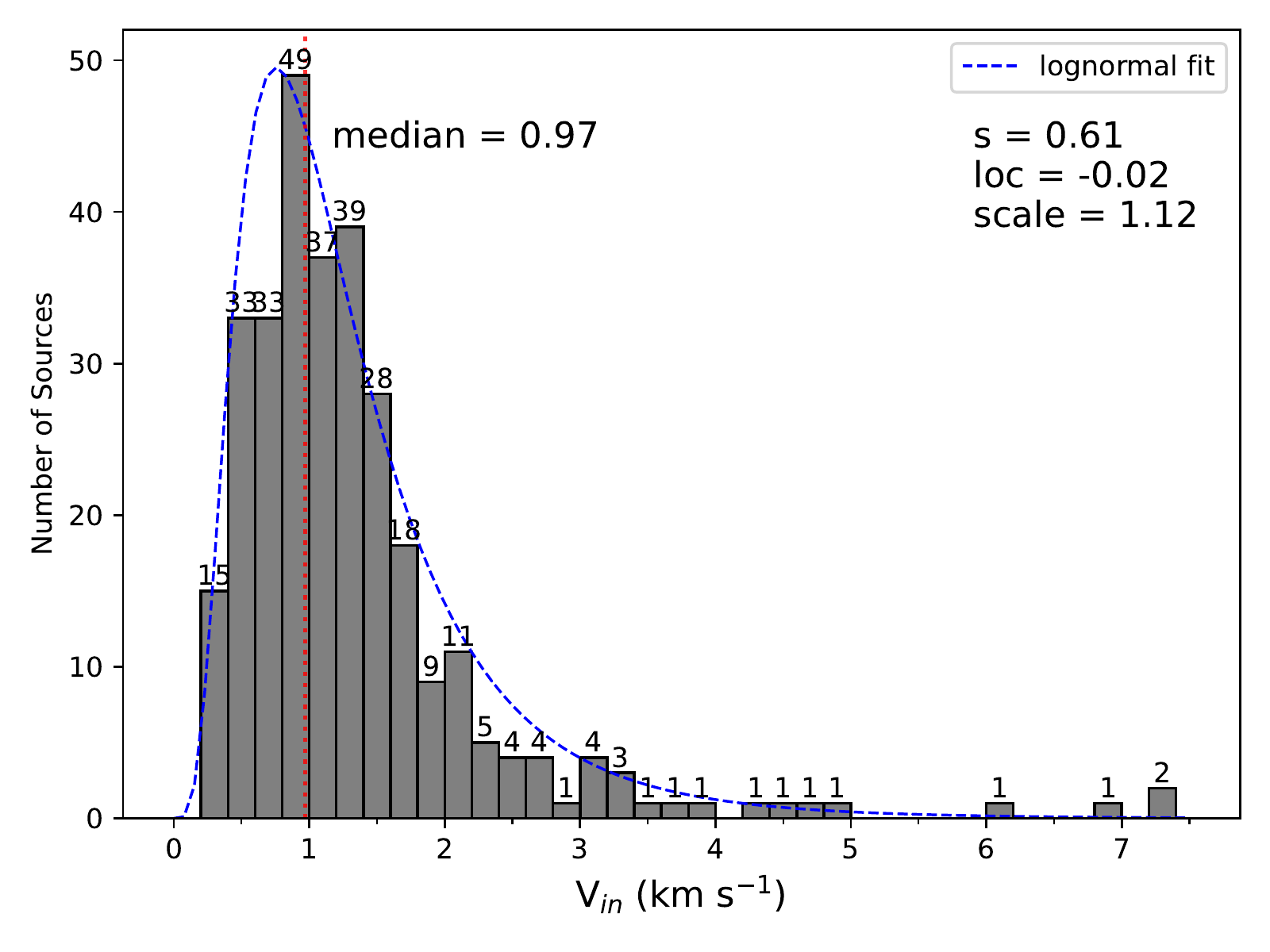}
      \caption{\small Distribution of infall velocity. The red\\dotted line indicates the median value. The \\blue dashed line is the fitting of lognormal.}
      \label{fig:Vin}
   \end{minipage}%
   \begin{minipage}[t]{0.595\linewidth}
      \centering
      \includegraphics[width=80mm]{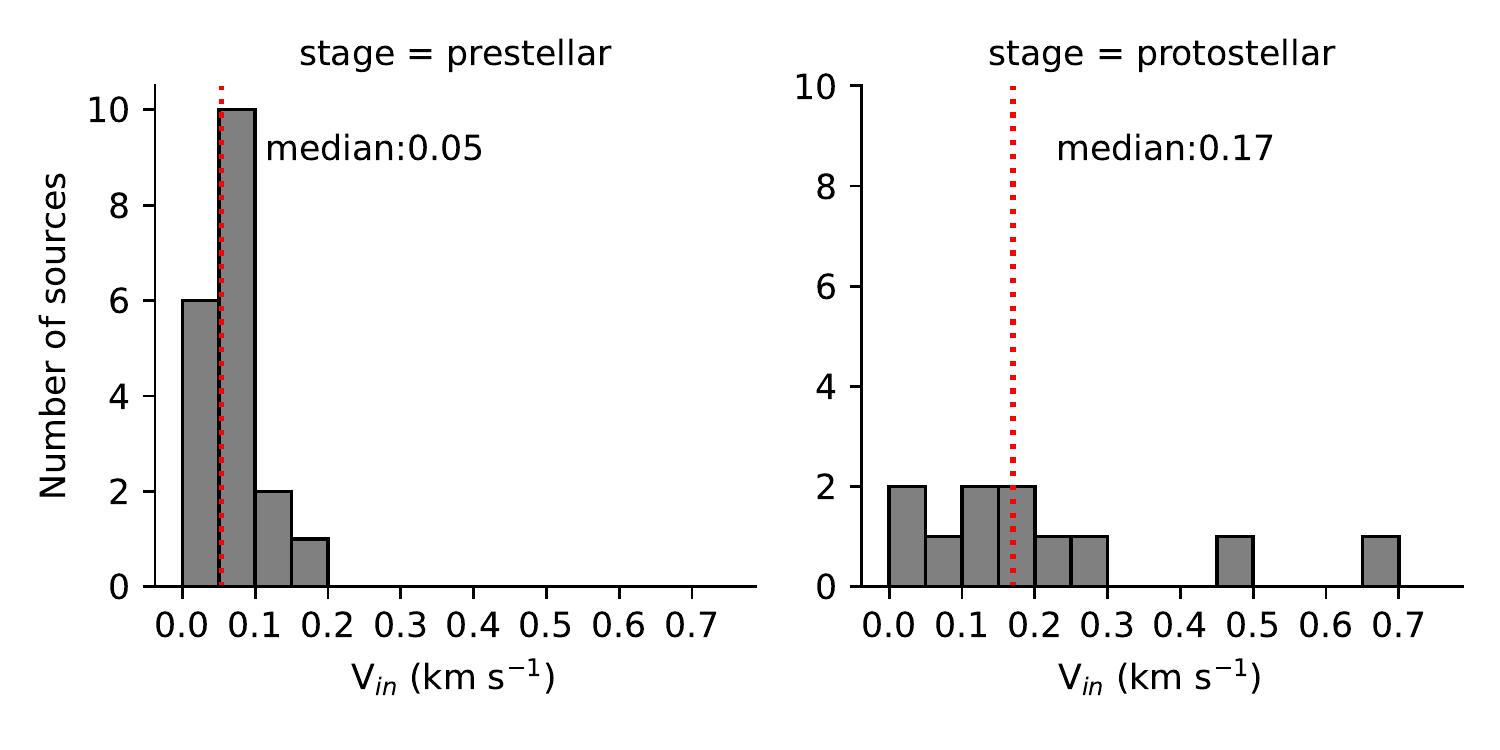}
      \caption{\small Distribution of infall velocity at different evolutionary stages for low-mass cores. The red dotted line represents the median value for each stage.}
      \label{fig:Vin_stages_low}
   \end{minipage}%
\end{figure}

Figure~\ref{fig:Vin} shows the distribution of infall velocities for 354 sources provided in the literature. We give the fitting of lognormal distribution (blue dashed line).
The parameters are marked in the upper right of the figure. Kolmogorov–Smirnov (KS) test shows the p value is 0.31. The median value (red dotted line) is 0.97 km s$^{-1}$.
The infall velocity varies with tracers and position across each core (\citealt{Keown2016}). \citet{Liu2013} believes that infall velocity may be related to the stage of evolution. In order to further study this, we conduct research according to different type of mass and stages.
Figure~\ref{fig:Vin_stages_low} shows the distribution of infall velocity to different evolutionary stages of low-mass cores.
The $V_{\text{in}}$ of almost all low-mass sources are less than 0.5 km s$^{-1}$. This is rather different from the result of \citet{Liu2013}, which shows the typical infall velocities in low-mass star-forming regions are $\sim$0.5 km s$^{-1}$.
The median values of prestellar and protostellar stage are 0.05 ($\pm$ 0.04) and 0.17 ($\pm$ 0.20) km s$^{-1}$, suggest that $V_{\text{in}}$ may not change with evolution for low-mass cores. However, this result should be interpreted with caution and await further investigation.
KS test assumes that the distribution of these two stages is consistent, and the P value is 0.013, rejecting the original hypothesis.

\begin{figure}[htbp]
   \centering
   \includegraphics[width=15cm, angle=0]{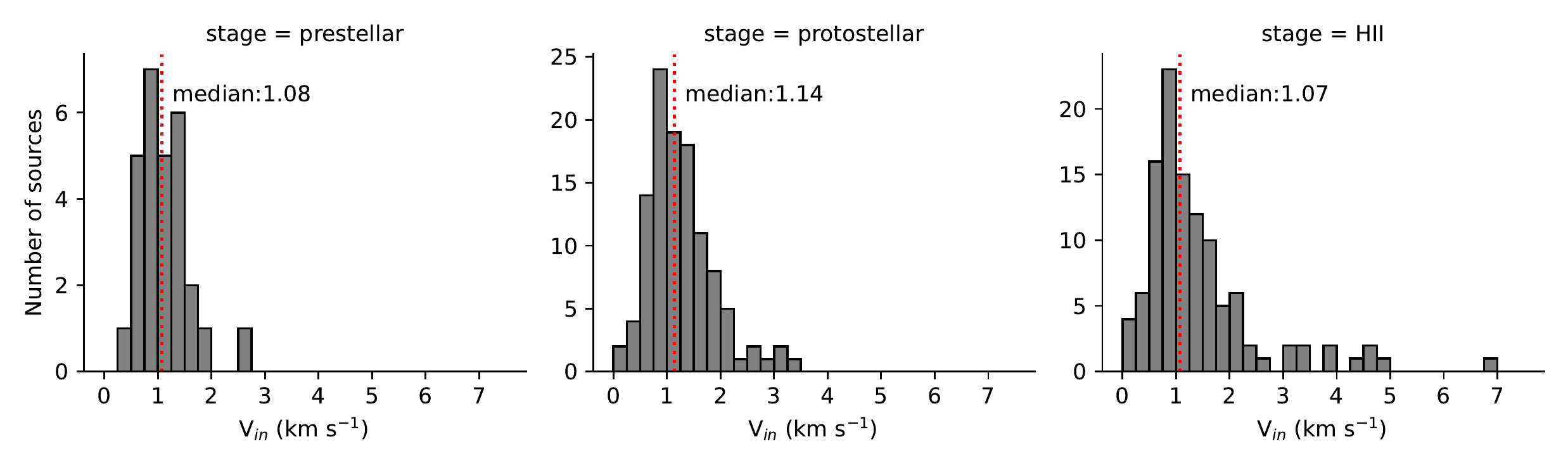}
   % \begin{minipage}[]{85mm}
   \caption{\small Distribution of infall velocity for high-mass clumps at different evolutionary stages. The red dotted line represents the median value of $V_{\text{in}}$ for each stage.}
   %\end{minipage}
   \label{fig:Vin_stages_high}
\end{figure}

Figure~\ref{fig:Vin_stages_high} shows the distribution of infall velocities for high-mass clumps at different stages. The median values of prestellar, protostellar and {H\small \hspace{0.1em}II} region are 1.08 ($\pm$ 0.49), 1.14 ($\pm$ 0.62) and 1.07 ($\pm$ 1.11) km s$^{-1}$, respectively, implying that the infall velocities of the high-mass clumps have a similar trend to that of the low-mass cores, that is, it may not change with evolutionary stages.
It can be seen from the figure that the distributions of these three stages are roughly the same. To confirm this, we performed the KS test, assuming the same distribution. The P value for the first two stages is 0.72, and for the stage of protostellar and {H\small \hspace{0.1em}II} is P = 0.59, which show that the original hypothesis is not rejected.
Skewness is a metrics of the distribution and extent of statistical data, defined as $\operatorname{skew}(X)=E\left[\left(\frac{X-\mu}{\sigma}\right)^{3}\right]$, which can characterize distribution asymmetry.
The skewness of the three stages of the massive infall clumps are 1.10, 1.10, 2.10, respectively, which indicates that the distribution of $V_{\text{in}}$ tends to be symmetrical at the first two evolutionary stages, while it tends to be rightwing at the {H\small \hspace{0.1em}II} region stage.

We have also analyzed the influence of different spectral lines on the infall velocity of the sources. The lines adopted by at least 15 sources are used for the following statistics. They are: HCO$^+$(1-0), CO (4-3), CS (2-1). 
For the cores with valid $V_{\text{in}}$ values, only CS (2-1) meets the statistical requirement. The median value is 0.05 ($\pm$ 0.04) km s$^{-1}$, which is basically consistent with the median value 0.07 km s$^{-1}$ of all low-mass cores. This result may change as the sample number of low-mass cores increases. 
For high-mass clumps, The median values estimated by HCO$^+$(1-0) and CO (4-3) are 1.03 ($\pm$ 0.85), 3.21($\pm$ 1.45) km s$^{-1}$, respectively. The number of $V_{\text{in}}$ estimated by HCO$^+$(1-0) accounts for more than 90\%, and is thus close to the median value 1.06 of all the high-mass clumps. 
The sources using CO (4-3) to derive $V_{\text{in}}$ are all from \citet*{Yue2021}.

On the other hand, \citet*{Rygl2013} calculated the source (G014.63-00.57 MM1) and showed that the $V_{\text{in}}$ derived from HCO$^+$(1-0) is about twice as large as that derived from HCO$^+$(4-3), but he thought the value derived from HCO$^+$(1-0) was overestimated. 
\citet*{Saral2018} used two lines (HCO$^+$ and HNC) to estimate infall velocities of the same three sources, the results were similar taking into account the error. 
\citet*{Tang2019} estimated $V_{\text{in}}$ of one source (G192.16-3.84) with C$^{18}$O (2-1) and HCO$^+$(3-2), indicating that the result derived from HCO$^+$(3-2) are slightly larger than C$^{18}$O (2-1). 
It can be seen that, due to the lack of total samples, we are temporarily unable to systematically study the influence of different spectral lines on the infall velocity.

\subsection{Mass Infall Rate}

The mass infall rate can be estimated by
\begin{equation}
   \dot{M}_{\mathrm{in}}=4 \pi R^{2} \mu m_{\mathrm{H}} n_{\mathrm{H_2}} V_{\mathrm{in}}
\end{equation}
where $\mu=2.8$ (\citealt{Wang2014}) is the mean molecular weight, $m_{\text{H}}$ is the mass of the hydrogen atom, $R$ is the radius of source (here the source is assumed to be spherical), V$_{\text{in}}$ is the infall velocity, $n_{\mathrm{H_2}}$ is the mean volume density, which has a conversion relationship with the column density $N_{\mathrm{H_2}}$:
\begin{equation}
   n_{\mathrm{H_2}}=\frac{3 N_{\mathrm{H_2}}}{R}
\end{equation}

\begin{figure}[htbp]
   \centering
   \includegraphics[width=8cm, angle=0]{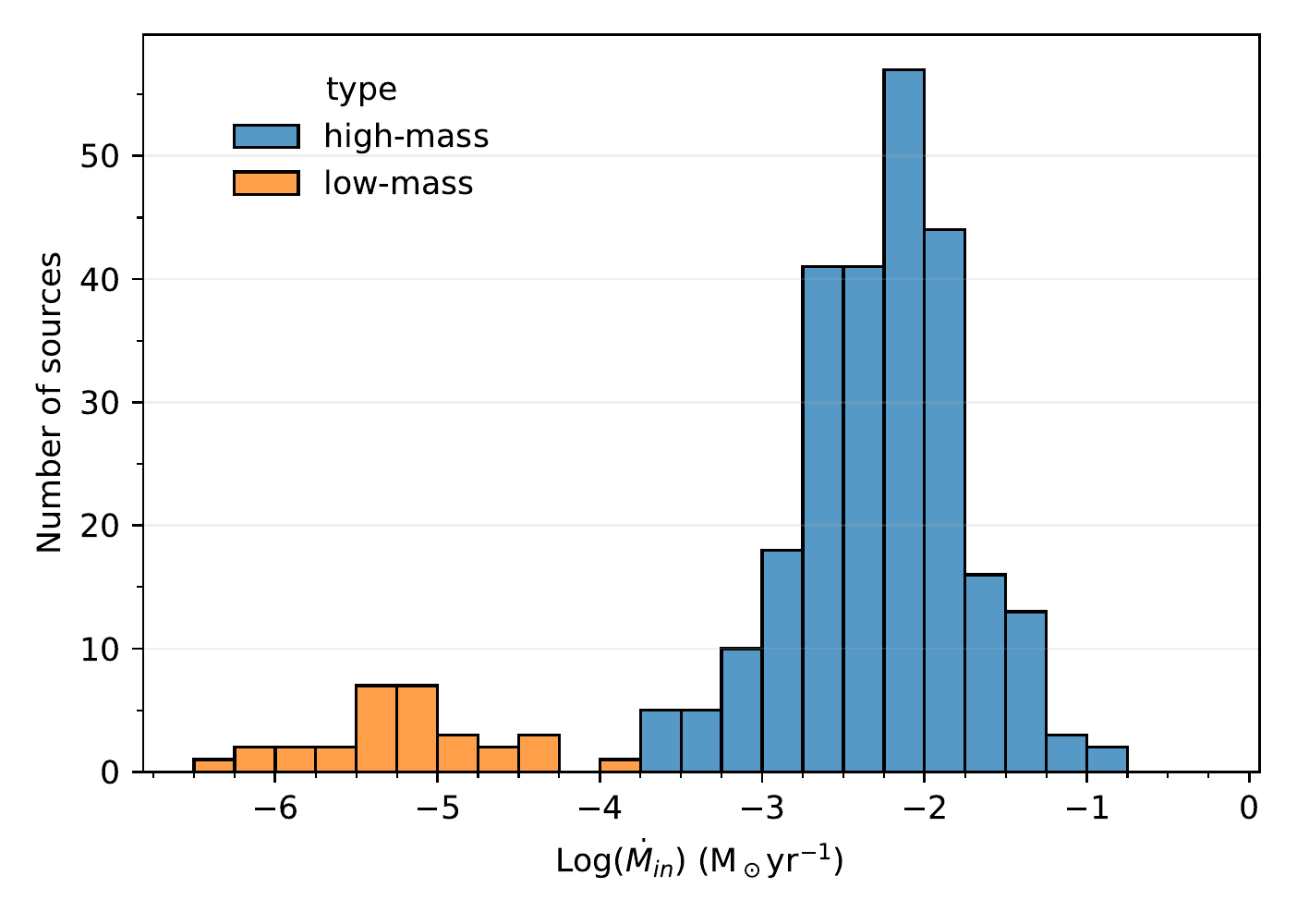}
   % \begin{minipage}[]{85mm}
   \caption{\small Distribution of mass infall rate provided in the literature.}
   %\end{minipage}
   \label{fig:log_mass_rate_type}
\end{figure}

Figure~\ref{fig:log_mass_rate_type} shows the distribution of mass infall rate of the sources provided in the literature. We could see from the figure that the distribution can be divided into two part, this separation comes from the different type of mass. Among them, we do not know the distribution of low-mass cores because of fewer samples, but for high-mass clumps, it presents an evident lognormal distribution with the P value of 0.82 of KS test.
% Figure~\ref{fig:log_mass_rate_type} shows the distribution of the mass infall rate of high-mass and low-mass.
The range of mass infall rate for low-mass cores are 10$^{-7}$ -- 10$^{-4}$ M$_{\odot} \text{yr}^{-1}$, and almost all high-mass clumps are 10$^{-4}$ -- 10$^{-1}$ M$_{\odot} \text{yr}^{-1}$ with only one exception.
This is consistent to previous work in high-mass star-forming regions range between 10$^{-4}$ -- 10$^{-2}$ M$_{\odot} \text{yr}^{-1}$ (e.g., \citealt{Wu2009,Liu2013,Qiu2012,Qin2016}), which are far larger than that of low-mass sources (\citealt{Beuther2005,Padoan2014,Kim2021}).

\begin{figure}[htbp]
   \centering
   \includegraphics[width=15cm, angle=0]{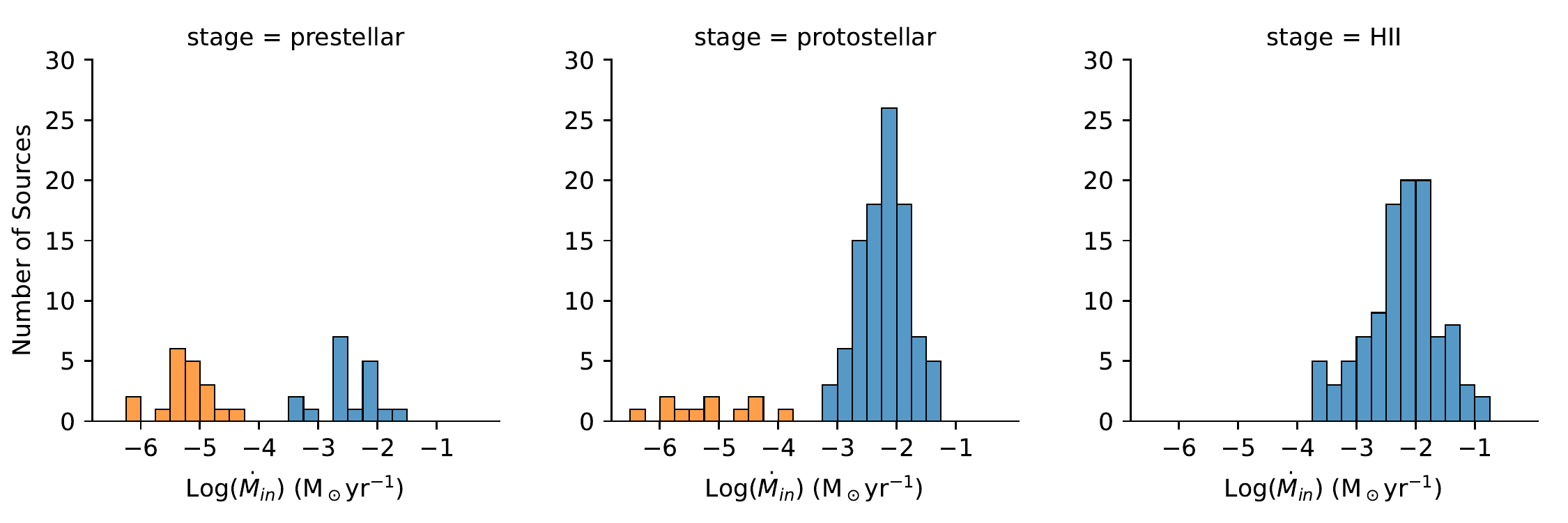}
   % \begin{minipage}[]{85mm}
   \caption{\small Distribution of mass infall rate for sources at different evolutionary stages. Orange represents low-mass cores and blue represents high-mass clumps.}
   %\end{minipage}
   \label{fig:mass_infall_rate_stages}
\end{figure}

\citet{Padoan2014} proposed that in most cases, the $\dot{M}_{\text{in}}$ is initially the order of 10$^{-5}$ M$_{\odot} \text{yr}^{-1}$, and may decay rapidly when forming low-mass stars, or maintain relatively large when forming massive stars. \citet{Yue2021} concluded that mass infall rate is independent of the evolutionary stage.
Figure~\ref{fig:mass_infall_rate_stages} shows the distribution of mass infall rate for sources at different evolutionary stages. The orange represents low-mass cores and the blue represents high-mass clumps.
The median values of $\dot{M}_{\mathrm{in}}$ at each stage for low-mass cores are 6.03 ($\pm$ 0.41) $\times$ 10$^{-6}$ and 6.46 ($\pm$ 0.77) $\times$ 10$^{-6}$ M$_{\odot} \text{yr}^{-1}$ respectively, and for high-mass clumps are 2.63 ($\pm$ 0.46) $\times$ 10$^{-3}$, 6.76 ($\pm$ 0.41) $\times$ 10$^{-3}$ , 7.24 ($\pm$ 0.61) $\times$ 10$^{-3}$ M$_{\odot} \text{yr}^{-1}$ respectively. Each of them is in the same order of magnitude. We do not find that the mass infall rates of either high-mass or low-mass sources change significantly with evolutionary stages.
Note that due to the small number of samples, especially low-mass samples, we need more research to discuss the trend. In addition, the error of distance may also have some influence on the statistical results.

% \begin{figure}[htbp]
%    \begin{minipage}[t]{0.595\linewidth}
%       \centering
%       \includegraphics[width=75mm]{picture/mass_infall_rate.pdf}
%       \caption{\small Distribution of mass infall rate for sources provided \\ in the literature.}
%       \label{fig:mass_infall_rate}
%    \end{minipage}%
%    \begin{minipage}[t]{0.495\linewidth}
%       \centering
%       \includegraphics[width=65mm]{picture/log_mass_rate_type.pdf}
%       \caption{\small Distribution of mass infall rate for high-mass and low-mass sources.}
%       \label{fig:log_mass_rate_type}
%    \end{minipage}%
% \end{figure}

\subsection{Association with Masers and Outflows}

Infall motion is a part of the initial activity of star formation, and maser is also a common phenomenon in star-forming regions. Class I CH$_3$OH masers are usually associated with outflow because of the collisional pumping mechanism, while class II CH$_3$OH masers generally occur in compact regions and possibly associated with massive star-forming regions due to the mechanism of radiative pumping. In order to study the correlation, we matched the summarized infall sources with the existing maser database (https://maserdb.net).

If a master source is found within $1'$ of the infall source, we think that they are associated. Table~\ref{Tab:maser} give the association rate between sources and masers with different type of mass.
There are 208 sources being associated with masers, the association rate is 46\%, and the number for high-mass and low-mass sources are 190 of 352 (54\%) and 10 of 54 (19\%), respectively.
The number associated with masers of Class I CH$_3$OH, Class II CH$_3$OH, H$_2$O, and OH are 130 (29\%), 122 (27\%), 165 (36\%), 75 (16\%), respectively.
We find that the association rates between high-mass sources and the four kind of masers are generally greater than those of low-mass. This may indicate the star formation activity near massive sources is more intense.
In addition, we can see that there's no low-mass sources associate with class II CH$_3$OH maser, which is consistent with the conclusion that the class II CH$_3$OH maser is only associated with the massive star-forming region.
Of the four masers, the association rate of H$_2$O is slightly higher, and that of OH is lower for both high-mass and low-mass sources. This may be related to the fact that the samples are not unbiased.

In the early stages of star formation, collapse and infall motion are often accompanied by molecular outflows (\citealt{Li2014}). The study of the relationship between infall and outflow is helpful to better understand the activity mechanism of star formation.
Therefore we also matched these sources with outflows provided in some literature (\citealt{Wu2004,Maud2015,Li2019a,Zhang2020}) in the range of $2'$. There are 52 (11\%) sources being associated with outflows, which is shown in the appendix table~\ref{Tab:Catalog}. Part of the reason for the low association rate may be that the outflow samples are incomplete.

% outflow

% \usepackage{multirow}
\begin{table}[]
   \caption[]{The maser association rate of sources with different masses.}\label{Tab:maser}
   \begin{center}
      \begin{tabular}{llcccccccccc}
         \hline
         \multicolumn{2}{c}{\textbf{}}   &
         \multicolumn{2}{c}{CH$_3$OH-I}  &
         \multicolumn{2}{c}{CH$_3$OH-II} &
         \multicolumn{2}{c}{H$_2$O}      &
         \multicolumn{2}{c}{OH}          &
         \multicolumn{2}{c}{maser$^*$}                                                                                     \\
         \hline
         \multirow{4}{*}{\begin{tabular}[c]{@{}l@{}}infall\\ sources\end{tabular}}
                                         & high-mass (352) & 121 & 34\% & 116 & 33\% & 151 & 43\% & 70 & 20\% & 190 & 54\% \\
                                         & low-mass (54)   & 7   & 13\% & 0   & 0\%  & 8   & 15\% & 2  & 4\%  & 10  & 19\% \\
                                         & unknown (50)    & 2   & 4\%  & 6   & 12\% & 6   & 12\% & 3  & 6\%  & 8   & 16\% \\
                                         & total (456)     & 130 & 29\% & 122 & 27\% & 165 & 36\% & 75 & 16\% & 208 & 46\% \\
         \hline
      \end{tabular}
   \end{center}
   \tablecomments{0.95\textwidth}{* idicates the number of masers detected in any of the four kind of masers.}
\end{table}

% % 此方式是把多张图并排放在一起,因此图号会随着minipage的数量依次增加
% \begin{figure}[h]
%    \begin{minipage}[t]{0.495\linewidth}
%    \centering
%     \includegraphics[width=60mm]{data/image/T_ex.eps}
%     \caption{{\small Amplitudes evolution of . . .} }
%    \end{minipage}%
%    \begin{minipage}[t]{0.495\linewidth}
%    \centering
%     \includegraphics[width=60mm]{data/image/log_N.eps}
%    \caption{{\small Amplitude variation of AN Lyn.}}
%    \end{minipage}%
%    \label{fig1}
%  \end{figure}

% 此方式是一张图里面放多个子图,因此图号只增加1.
%  \begin{figure}[htbp]  %[htbp]中的h是浮动的意思，H表示固定
%    \centering    %居中

%    \subfloat[] %第一张子图，[]内可填写子图的标注
%    {
%        \begin{minipage}[t]{0.495\linewidth} %子图以0.33倍行宽显示
%            \centering          %子图居中
%            \includegraphics[width=1\linewidth]{data/image/T_ex.eps}   %子图空间内以行宽的0.5倍大小显示
%            \captionsetup{font={scriptsize}}
%        \end{minipage}
%    } % 注意这里不能回车空行，否则两张图会上下排列，而不是并排排列
%    \subfloat[] %第二张子图
%    {
%        \begin{minipage}[t]{0.495\linewidth} %子图以0.33倍行宽显示
%            \centering      %子图居中
%            \includegraphics[width=1\linewidth]{data/image/log_N.eps}   %子图空间内以行宽的0.5倍大小显示
%            \captionsetup{font={scriptsize}}
%        \end{minipage}
%    }%
%    \caption{(a):; (b): }  %整个大图的标注
%    \captionsetup{font={scriptsize}}
%    \label{fig1}  %图片引用标记
% \end{figure}

% \section{}           %% first-level sections will be auto-capitalized
% \label{sect:discussion}9

\section{Summary}           %% first-level sections will be auto-capitalized
\label{sect:Summary}

We searched the literature related to infall study since 1994, summarized infall sources and made some statistics on the physical properties. The result is summarized below.

1. A total of 456 sources have been cataloged, including 352 (77\%) sources with high mass, 55 (12\%) with low mass, and the remaining 49 (11\%) cannot be categorized. We further divide the high-mass sources into three groups according to the evolutionary stage, that is prestellar, protostellar, {H\small \hspace{0.1em}II} region. The number in the groups are 48 (11\%), 215 (47\%), 125 (27\%), respectively. The remaining 69 (15\%) sources have no evolutionary information. Most of the sources are located within 5 kpc from the sun.
We divide the sources into clumps and cores according to their sizes. Vast majority (97\%) of the high-mass sources are clumps, and most (85\%) of the low-mass sources are cores.
Although the current observations are far from complete, the present evidence shows that the star formation activity within the solar circle is more intense than the outer.
We summarized the optically thick lines utilized to identify the infall signatures. About 77\% of infall sources are identified with HCO$^+$ (1-0), suggested this line is more effective in identifying infall sources than the others.

2. The H$_2$ column densities of high-mass clumps ranges between 10$^{22}$ and 10$^{24}$ cm$^{-2}$, and those of low-mass cores are about 10$^{21}$ cm$^{-2}$.
The median value of high-mass clumps is 4.8 $\times$10$^{22}$ cm$^{-2}$, which is one order of magnitude higher than that of low-mass cores.
% The order of the H$_2$ column density for low-mass stars is 10$^{21}$ cm$^{-2}$, and for high-mass stars is about 10$^{22}$ cm$^{-2}$ $\sim$ 10$^{24}$ cm$^{-2}$. 

3. We counted the skewness of sources and classified them according to different mass and evolutionary stages, but it does not show significant differences with the mass and evolutionary stage.
% The median value of all sources for the skewness parameter $\delta V$  is -0.52, and for infall velocity is 0.98 km s$^{-1}$.

4. The median value of infall velocities ($V_{\text{in}}$) for high-mass clumps is 1.12 km s$^{-1}$ with the RMS 1.02, and the $V_{\text{in}}$ of nearly all low-mass cores are less than 0.5 km s$^{-1}$. There is no evidence that infall velocity changes with evolution.

5. The range of mass infall rates for low-mass cores are 10$^{-7}$ $\sim$ 10$^{-4}$ M$_{\odot} \text{yr}^{-1}$, and for high-mass clumps are 10$^{-4}$ $\sim$ 10$^{-1}$ M$_{\odot} \text{yr}^{-1}$ with one exception. We do not find that the mass infall rates change with evolutionary stages for either high-mass or low-mass sources.

6. There are 208 (46\%) sources associate with masers within an angular radius of $1'$. The number of infall sources being associated with masers of Class I CH$_3$OH, Class II CH$_3$OH, H$_2$O, OH are 130 (29\%), 122 (27\%), 165 (36\%), 75 (16\%). There are 52 (11\%) sources being associated with outflows within $2'$.

\begin{acknowledgements}

   This work is supported by the National Key R\&D Program of China (grant No. 2017YFA0402702), and the National Natural Science Foundation of China (NSFC, grant Nos. 11873093 and U2031202), Z. Chen acknowledges the support from the NSFC general grant 11903083.

\end{acknowledgements}

\appendix
\section{Table of All Collected Infall Sources}
\begin{landscape}
   % \begin{sidewaystable}                       % 旋转表格，与longtable不兼容,只可用于一页,不能跨页
   {\renewcommand{\arraystretch}{0.9}         % 设置行间距
      \small                           % 控制字体大小，tiny,scriptsize,footnotesize,small,normalsize,large,Large,LARGE,huge,HUGE...
      \setlength\LTleft{-0.05in}               % 设置左边距，-1为超出左边距-1
      \setlength\LTright{2in plus 1 fill}     % 设置右边距 (实践发现不起作用)
      \setlength{\tabcolsep}{2.3pt}           % 设置列间距,默认为6pt,可以利用列间距来间接控制右边距
\begin{longtable}{cccccccccccccccccl}
         % \begin{center}
         \caption[]{An Excerpt of the Physical Properties of Infall Sources.}\label{Tab:Catalog}                                                                                                                                                                                           \\

         %%Please Capitalize the First Letter of Each Notional Word in table's caption
         % \resizebox{1.05\textwidth}{!}{   % 自动调整表格宽度
         % \begin{tabular}{lllcccccccccccll}
         \hline\noalign{\smallskip}
         \hline\noalign{\smallskip}

         No.                                & R.A.                     & Dec.                     & Alias              & Dist.              & $V_{\text{LSR}}$           &
         $N$(H$_2$)                         & $n$(H$_2$)               & Mass                     & Lines              & $\delta v$         & $V_{in}$                   &
         $\dot M_{in}$                      & Type                     & Stage                    & Maser              & Outflow            & Ref.                                                                                                                                         \\
                                            & \scriptsize{(J2000)}     & \scriptsize{(J2000)}     &                    & \scriptsize{(kpc)} & \scriptsize{(km s$^{-1}$)} &
         \scriptsize{(cm$^{-2}$)}           & \scriptsize{(cm$^{-3}$)} & \scriptsize{(M$_\odot$)} &                    &                    & \scriptsize{(km s$^{-1}$)} &
         \scriptsize{(M$_\odot$ yr$^{-1}$)} &                          &                          &                    &                    &                                                                                                                                              \\
         \scriptsize{(1)}                   & \scriptsize{(2)}         & \scriptsize{(3)}         & \scriptsize{(4)}   & \scriptsize{(5)}   & \scriptsize{(6)}           &
         \scriptsize{(7)}                   & \scriptsize{(8)}         & \scriptsize{(9)}         & \scriptsize{(10)}  & \scriptsize{(11)}  & \scriptsize{(12)}          &
         \scriptsize{(13)}                  & \scriptsize{(14)}        & \scriptsize{(15)}        & \scriptsize{(16)}  & \scriptsize{(17)}  & \scriptsize{(18)}                                                                                                                            \\
         \hline
         \endhead % 使每一页重复出现表头
         \hline
         \endfoot % 使每一页底部重复出现水平线
         \hline
         \endlastfoot

         1                                  & 00:35:40.1               & +66:14:23                & G121.34+3.42       & 0.7                & -5.3                       & 9.8E+21 &        &           & \textbf{1},9         & -0.47 &      &        &      &   &      &   & 3           \\
         2                                  & 00:36:47.5               & +63:29:01                & G121.30+0.66       & 1.69               & -17.6                      &         &        &           & 2                    & -0.54 &      &        & high &   & 123  & Y & 19          \\
         3                                  & 00:36:53.6               & +63:28:03                & G121.31+0.64       & 1.7                & -17.3                      & 1.7E+22 &        &           & \textbf{1},9         & -0.63 & 0.29 & 1.6E-3 &      & 2 &      & Y & 3           \\
         4                                  & 00:52:25.1               & +56:33:54                & G123.07-6.31       & 2.53               & -30.4                      &         &        &           & 1,\textbf{2},10      & -0.20 &      &        & high & 3 & 123  & Y & 19,20       \\
         5                                  & 02:19:51.8               & +61:03:26                & G133.42+0.00       & 1.46               & -15.2                      & 4.8E+21 &        &           & 9                    & -0.59 &      &        &      & 2 &      &   & 3           \\
         6                                  & 02:27:03.8               & +61:52:25                & W3(OH)             & 1.96               & -48.0                      &         & 6.0E+6 &           & \textbf{1},10        & -0.23 & 0.06 & 3.0E-4 & high & 3 & 1234 & Y & 14,15,19,20 \\
\end{longtable}
   }
   % \vspace{}
   % \end{sidewaystable}
\end{landscape}

%%%%%%%%%%%%%%%% 参考文献 %%%%%%%%%%%%%%%%%%%

% \renewcommand\refname{参考文献}

\bibliographystyle{raa}
\bibliography{reference}

\begin{thebibliography}{88}
\providecommand\natexlab[1]{#1}
\providecommand\JournalTitle[1]{#1}

\bibitem[{Beltr{\'a}n} \& {de Wit}(2016)]{Beltran2016}
{Beltr{\'a}n}, M.~T., \& {de Wit}, W.~J. 2016, \aapr, 24, 6

\bibitem[Beuther {et~al.}(2005)]{Beuther2005}
Beuther, H., Sridharan, T.~K., \& Saito, M. 2005, The Astrophysical Journal,
  634, L185

\bibitem[Bonnell \& Bate(2002)]{Bonnell2002}
Bonnell, I.~A., \& Bate, M.~R. 2002, Monthly Notices of the Royal Astronomical
  Society, 336, 659

\bibitem[Bonnell {et~al.}(2004)]{Bonnell2004}
Bonnell, I.~A., Vine, S.~G., \& Bate, M.~R. 2004, Monthly Notices of the Royal
  Astronomical Society, 349, 735

\bibitem[{Cesaroni} {et~al.}(2007)]{Cesaroni2007}
{Cesaroni}, R., {Galli}, D., {Lodato}, G., {Walmsley}, C.~M., \& {Zhang}, Q.
  2007, in Protostars and Planets V, ed. B.~{Reipurth}, D.~{Jewitt}, \&
  K.~{Keil}, 197

\bibitem[{Churchwell} {et~al.}(2010)]{Churchwell2010}
{Churchwell}, E., {Sievers}, A., \& {Thum}, C. 2010, \aap, 513, A9

\bibitem[Contreras {et~al.}(2018)]{Contreras2018}
Contreras, Y., Sanhueza, P., Jackson, J.~M., {et~al.} 2018, The Astrophysical
  Journal, 861, 14

\bibitem[Cunningham {et~al.}(2018)]{Cunningham2018}
Cunningham, N., Lumsden, S.~L., Moore, T.~J., Maud, L.~T., \& Mendigut{\'{i}}a,
  I. 2018, Monthly Notices of the Royal Astronomical Society, 477, 2455

\bibitem[{De Vries} \& Myers(2005)]{DeVries2005}
{De Vries}, C.~H., \& Myers, P.~C. 2005, The Astrophysical Journal, 620, 800

\bibitem[{Di Francesco} {et~al.}(2001)]{DiFrancesco2001}
{Di Francesco}, J., Myers, P.~C., Wilner, D.~J., Ohashi, N., \& Mardones, D.
  2001, The Astrophysical Journal, 562, 770

\bibitem[{Djordjevic} {et~al.}(2019)]{Djordjevic2019}
{Djordjevic}, J.~O., {Thompson}, M.~A., {Urquhart}, J.~S., \& {Forbrich}, J.
  2019, \mnras, 487, 1057

\bibitem[Evans(1999)]{Evans1999}
Evans, N.~J. 1999, Annual Review of Astronomy and Astrophysics, 37, 311

\bibitem[Fa{\'{u}}ndez {et~al.}(2004)]{Faundez2004}
Fa{\'{u}}ndez, S., Bronfman, L., Garay, G., {et~al.} 2004, Astronomy and
  Astrophysics, 426, 97

\bibitem[Fuller {et~al.}(2005)]{Fuller2005}
Fuller, G.~A., Williams, S.~J., \& Sridharan, T.~K. 2005, Astronomy and
  Astrophysics, 442, 949

\bibitem[{Gregersen}(1998)]{Gregersen1998}
{Gregersen}, E.~M. 1998, {Collapse and Beyond: an Investigation of the Star
  Formation Process}, PhD thesis, THE UNIVERSITY OF TEXAS AT AUSTIN

\bibitem[Gregersen(2000)]{Gregersen2000}
Gregersen, E.~M. 2000, The Astrophysical Journal, 533, 440

\bibitem[Gregersen {et~al.}(1997)]{Gregersen1997}
Gregersen, E.~M., {Evans II}, N.~J., Zhou, S., \& Choi, M. 1997, The
  Astrophysical Journal, 484, 256

\bibitem[Guzm{\'{a}}n {et~al.}(2015)]{Guzman2015}
Guzm{\'{a}}n, A.~E., Sanhueza, P., Contreras, Y., {et~al.} 2015, Astrophysical
  Journal, 815, 130

\bibitem[He {et~al.}(2015)]{He2015}
He, Y.~X., Zhou, J.~J., Esimbek, J., {et~al.} 2015, Monthly Notices of the
  Royal Astronomical Society, 450, 1926

\bibitem[He {et~al.}(2016)]{He2016}
He, Y.~X., Zhou, J.~J., Esimbek, J., {et~al.} 2016, Monthly Notices of the
  Royal Astronomical Society, 461, 2288

\bibitem[Hogerheijde \& {Van Der Tak}(2000)]{Hogerheijde2000}
Hogerheijde, M.~R., \& {Van Der Tak}, F.~F. 2000, Astronomy and Astrophysics,
  362, 697

\bibitem[Keown {et~al.}(2016)]{Keown2016}
Keown, J., Schnee, S., Bourke, T.~L., {et~al.} 2016, The Astrophysical Journal,
  833, 97

\bibitem[Keto {et~al.}(2015)]{Keto2015}
Keto, E., Caselli, P., \& Rawlings, J. 2015, Monthly Notices of the Royal
  Astronomical Society, 446, 3731

\bibitem[Kim {et~al.}(2021)]{Kim2021}
Kim, M.-R., Lee, C.~W., Maheswar, G., Myers, P.~C., \& Kim, G. 2021, The
  Astrophysical Journal, 910, 112

\bibitem[Klaassen {et~al.}(2012)]{Klaassen2012}
Klaassen, P.~D., Testi, L., \& Beuther, H. 2012, Astronomy and Astrophysics,
  538, A140

\bibitem[Klaassen \& Wilson(2007)]{Klaassen2007}
Klaassen, P.~D., \& Wilson, C.~D. 2007, The Astrophysical Journal, 663, 1092

\bibitem[Klaassen \& Wilson(2008)]{Klaassen2008}
Klaassen, P.~D., \& Wilson, C.~D. 2008, The Astrophysical Journal, 684, 1273

\bibitem[Klassen {et~al.}(2016)]{Klassen2016}
Klassen, M., Pudritz, R.~E., Kuiper, R., Peters, T., \& Banerjee, R. 2016, The
  Astrophysical Journal, 823, 28

\bibitem[Kurtz {et~al.}(1994)]{Kurtz1994}
Kurtz, S., Churchwell, E., \& Wood, D. O.~S. 1994, The Astrophysical Journal
  Supplement Series, 91, 659

\bibitem[Lee \& Myers(2011)]{Lee2011}
Lee, C.~W., \& Myers, P.~C. 2011, The Astrophysical Journal, 734, 60

\bibitem[Lee {et~al.}(1999)]{Lee1999}
Lee, C.~W., Myers, P.~C., \& Tafalla, M. 1999, The Astrophysical Journal, 526,
  788

\bibitem[Lee {et~al.}(2001)]{Lee2001}
Lee, C.~W., Myers, P.~C., \& Tafalla, M. 2001, The Astrophysical Journal
  Supplement Series, 136, 703

\bibitem[Leung \& Brown(1977)]{Leung1977}
Leung, C.~M., \& Brown, R.~L. 1977, The Astrophysical Journal, 214, L73

\bibitem[Li {et~al.}(2019)]{Li2019a}
Li, Y., Xu, Y., Sun, Y., {et~al.} 2019, The Astrophysical Journal Supplement
  Series, 242, 19

\bibitem[{Li} {et~al.}(2014)]{Li2014}
{Li}, Z.~Y., {Banerjee}, R., {Pudritz}, R.~E., {et~al.} 2014, in Protostars and
  Planets VI, ed. H.~{Beuther}, R.~S. {Klessen}, C.~P. {Dullemond}, \&
  T.~{Henning}, 173

\bibitem[Liu {et~al.}(2020)]{Liu2020}
Liu, S.-Y., Su, Y.-N., Zinchenko, I., {et~al.} 2020, The Astrophysical Journal,
  904, 181

\bibitem[Liu {et~al.}(2013)]{Liu2013}
Liu, T., Wu, Y., \& Zhang, H. 2013, The Astrophysical Journal, 776, 29

\bibitem[Liu {et~al.}(2011)]{Liu2011}
Liu, T., Wu, Y., Zhang, Q., {et~al.} 2011, The Astrophysical Journal, 728

\bibitem[Mardones {et~al.}(1997)]{Mardones1997}
Mardones, D., Myers, P.~C., Tafalla, M., {et~al.} 1997, The Astrophysical
  Journal, 489, 719

\bibitem[Maud {et~al.}(2015)]{Maud2015}
Maud, L.~T., Moore, T.~J., Lumsden, S.~L., {et~al.} 2015, Monthly Notices of
  the Royal Astronomical Society, 453, 645

\bibitem[McKee \& Tan(2002)]{McKee2002}
McKee, C.~F., \& Tan, J.~C. 2002, Nature, 416, 59

\bibitem[McKee \& Tan(2003)]{McKee2003}
McKee, C.~F., \& Tan, J.~C. 2003, The Astrophysical Journal, 585, 850

\bibitem[Motte {et~al.}(2018)]{Motte2018}
Motte, F., Bontemps, S., \& Louvet, F. 2018, Annual Review of Astronomy and
  Astrophysics, 56, 41

\bibitem[Myers {et~al.}(1996)]{Myers1996}
Myers, P.~C., Mardones, D., Tafalla, M., Williams, J.~P., \& Wilner, D.~J.
  1996, The Astrophysical Journal, 465, L133

\bibitem[Padoan {et~al.}(2014)]{Padoan2014}
Padoan, P., Haugb{\o}lle, T., \& Nordlund, {\AA}. 2014, The Astrophysical
  Journal, 797, 32

\bibitem[Peretto {et~al.}(2013)]{Peretto2013}
Peretto, N., Fuller, G.~A., Duarte-Cabral, A., {et~al.} 2013, Astronomy and
  Astrophysics, 555, A112

\bibitem[Qin {et~al.}(2016)]{Qin2016}
Qin, S.~L., Schilke, P., Wu, J., {et~al.} 2016, Monthly Notices of the Royal
  Astronomical Society, 456, 2681

\bibitem[Qiu {et~al.}(2012)]{Qiu2012}
Qiu, K., Zhang, Q., Beuther, H., \& Fallscheer, C. 2012, The Astrophysical
  Journal, 756, 170

\bibitem[{Redman} {et~al.}(2008)]{Redman2008}
{Redman}, M.~P., {Khanzadyan}, T., {Loughnane}, R.~M., \& {Carolan}, P.~B.
  2008, in Astronomical Society of the Pacific Conference Series, Vol. 387,
  Massive Star Formation: Observations Confront Theory, ed. H.~{Beuther},
  H.~{Linz}, \& T.~{Henning}, 38

\bibitem[Reid {et~al.}(2016)]{Reid2016}
Reid, M.~J., Dame, T.~M., Menten, K.~M., \& Brunthaler, A. 2016, The
  Astrophysical Journal, 823, 77

\bibitem[Reid {et~al.}(2019)]{Reid2019}
Reid, M.~J., Menten, K.~M., Brunthaler, A., {et~al.} 2019, The Astrophysical
  Journal, 885, 131

\bibitem[Reiter {et~al.}(2011)]{Reiter2011}
Reiter, M., Shirley, Y.~L., Wu, J., {et~al.} 2011, The Astrophysical Journal,
  740, 1

\bibitem[Rygl {et~al.}(2013)]{Rygl2013}
Rygl, K. L.~J., Wyrowski, F., Schuller, F., \& Menten, K.~M. 2013, Astronomy
  and Astrophysics, 549, A5

\bibitem[Saral {et~al.}(2018)]{Saral2018}
Saral, G., Audard, M., \& Wang, Y. 2018, Astronomy and Astrophysics, 620, A158

\bibitem[{Sarrasin} {et~al.}(2010)]{Sarrasin2010}
{Sarrasin}, E., {Abdallah}, D.~B., {Wernli}, M., {et~al.} 2010, \mnras, 404,
  518

\bibitem[Schnee {et~al.}(2013)]{Schnee2013}
Schnee, S., Brunetti, N., {Di Francesco}, J., {et~al.} 2013, The Astrophysical
  Journal, 777, 121

\bibitem[Sch{\"{o}}ier {et~al.}(2005)]{Schoier2005}
Sch{\"{o}}ier, F.~L., {Van Der Tak}, F.~F., {Van Dishoeck}, E.~F., \& Black,
  J.~H. 2005, Astronomy and Astrophysics, 432, 369

\bibitem[Shirley {et~al.}(2003)]{Shirley2003}
Shirley, Y.~L., {Evans II}, N.~J., Young, K.~E., Knez, C., \& Jaffe, D.~T.
  2003, The Astrophysical Journal Supplement Series, 149, 375

\bibitem[Shu(1987)]{Shu1987}
Shu, F. 1987, Annual Review of Astronomy and Astrophysics, 25, 23

\bibitem[Smith {et~al.}(2013)]{Smith2013}
Smith, R.~J., Shetty, R., Beuther, H., Klessen, R.~S., \& Bonnell, I.~A. 2013,
  The Astrophysical Journal, 771, 24

\bibitem[Sohn {et~al.}(2007)]{Sohn2007}
Sohn, J., Lee, C.~W., Park, Y., {et~al.} 2007, The Astrophysical Journal, 664,
  928

\bibitem[Sridharan {et~al.}(2002)]{Sridharan2002}
Sridharan, T.~K., Beuther, H., Schilke, P., Menten, K.~M., \& Wyrowski, F.
  2002, The Astrophysical Journal, 566, 931

\bibitem[Storm {et~al.}(2016)]{Storm2016}
Storm, S., Mundy, L.~G., Lee, K.~I., {et~al.} 2016, The Astrophysical Journal,
  830, 127

\bibitem[Su {et~al.}(2019)]{Su2019}
Su, Y.-N., Liu, S.-Y., Li, Z.-Y., {et~al.} 2019, The Astrophysical Journal,
  885, 98

\bibitem[Su {et~al.}(2021)]{Su2021}
Su, Y., Yang, J., Yan, Q.-Z., {et~al.} 2021, The Astrophysical Journal, 910,
  131

\bibitem[Sun \& Gao(2009)]{Sun2009}
Sun, Y., \& Gao, Y. 2009, Monthly Notices of the Royal Astronomical Society,
  392, 170

\bibitem[{Tan} {et~al.}(2014)]{Tan2014}
{Tan}, J.~C., {Beltr{\'a}n}, M.~T., {Caselli}, P., {et~al.} 2014, in Protostars
  and Planets VI, ed. H.~{Beuther}, R.~S. {Klessen}, C.~P. {Dullemond}, \&
  T.~{Henning}, 149

\bibitem[Tang {et~al.}(2019)]{Tang2019}
Tang, M.~Y., Qin, S.~L., Liu, T., \& Wu, Y.~F. 2019, Research in Astronomy and
  Astrophysics, 19, 040

\bibitem[Tsamis {et~al.}(2008)]{Tsamis2008}
Tsamis, Y.~G., Rawlings, J.~M., Yates, J.~A., \& Viti, S. 2008, Monthly Notices
  of the Royal Astronomical Society, 388, 898

\bibitem[{Vasyunina} {et~al.}(2011)]{Vasyunina2011}
{Vasyunina}, T., {Linz}, H., {Henning}, T., {et~al.} 2011, \aap, 527, A88

\bibitem[Wang {et~al.}(2014)]{Wang2014}
Wang, K., Zhang, Q., Testi, L., {et~al.} 2014, Monthly Notices of the Royal
  Astronomical Society, 439, 3275

\bibitem[Wu \& {Evans II}(2003)]{Wu2003}
Wu, J., \& {Evans II}, N.~J. 2003, The Astrophysical Journal, 592, L79

\bibitem[Wu {et~al.}(2009)]{Wu2009}
Wu, Y., Qin, S.~L., Guan, X., {et~al.} 2009, Astrophysical Journal, 697, 116

\bibitem[Wu {et~al.}(2004)]{Wu2004}
Wu, Y., Wei, Y., Zhao, M., {et~al.} 2004, Astronomy and Astrophysics, 426, 503

\bibitem[Wu {et~al.}(2005)]{Wu2005}
Wu, Y., Zhu, M., Wei, Y., {et~al.} 2005, The Astrophysical Journal, 628, L57

\bibitem[Xu {et~al.}(2016)]{Xu2016}
Xu, Y., Reid, M., Dame, T., {et~al.} 2016, Science Advances, 2, e1600878

\bibitem[Yang {et~al.}(2020{\natexlab{a}})]{Yang2020b}
Yang, Y., Jiang, Z.-B., Chen, Z.-W., {et~al.} 2020{\natexlab{a}}, Research in
  Astronomy and Astrophysics, 20, 115

\bibitem[Yang {et~al.}(2020{\natexlab{b}})]{Yang2020}
Yang, Y.~L., Evans, N.~J., Smith, A., {et~al.} 2020{\natexlab{b}}, The
  Astrophysical Journal, 891, 61

\bibitem[{Yorke} \& {Sonnhalter}(2002)]{Yorke2002}
{Yorke}, H.~W., \& {Sonnhalter}, C. 2002, \apj, 569, 846

\bibitem[Yuan {et~al.}(2018)]{Yuan2018}
Yuan, J., Li, J.-Z., Wu, Y., {et~al.} 2018, The Astrophysical Journal, 852, 12

\bibitem[Yue {et~al.}(2021)]{Yue2021}
Yue, Y.~H., Qin, S.~L., Liu, T., {et~al.} 2021, Research in Astronomy and
  Astrophysics, 21, 14

\bibitem[Zhang {et~al.}(2018)]{Zhang2018}
Zhang, G.~Y., Xu, J.~L., Vasyunin, A.~I., {et~al.} 2018, Astronomy and
  Astrophysics, 620, A163

\bibitem[Zhang {et~al.}(2007)]{Zhang2007}
Zhang, Q., Sridharan, T.~K., Hunter, T.~R., {et~al.} 2007, Astronomy and
  Astrophysics, 470, 269

\bibitem[Zhang {et~al.}(2020)]{Zhang2020}
Zhang, S., Yang, J., Xu, Y., {et~al.} 2020, The Astrophysical Journal
  Supplement Series, 248, 15

\bibitem[Zhou(1992)]{Zhou1992}
Zhou, S. 1992, The Astrophysical Journal, 394, 204

\bibitem[{Zhou} {et~al.}(1994)]{Zhou1994}
{Zhou}, S., {Evans}, Neal~J., I., {Koempe}, C., \& {Walmsley}, C.~M. 1994,
  \apj, 421, 854

\bibitem[Zhou {et~al.}(1993)]{Zhou1993}
Zhou, S., {Evans II}, N.~J., Koempe, C., \& Walmsley, C.~M. 1993, The
  Astrophysical Journal, 404, 232

\bibitem[{Zinnecker} \& {Yorke}(2007)]{Zinnecker2007}
{Zinnecker}, H., \& {Yorke}, H.~W. 2007, \araa, 45, 481

\end{thebibliography}

\end{document}